\pdfoutput=1

\documentclass[11pt,twoside,a4paper,cmspaper,final,collab]{cms-tdr}
\def\svnVersion{81107:81122}\def\cmsCernNo{2011-125}\def\cmsCernDate{\today}\def\cmsMessage{Submitted to the Journal of High Energy Physics}

\begin{document}\cmsNoteHeader{EWK-10-012}

\hyphenation{had-ron-i-za-tion}
\hyphenation{cal-or-i-me-ter}
\hyphenation{de-vices}
\RCS$Revision: 81122 $
\RCS$HeadURL: svn+ssh://alverson@svn.cern.ch/reps/tdr2/papers/EWK-10-012/trunk/EWK-10-012.tex $
\RCS$Id: EWK-10-012.tex 81122 2011-10-14 14:33:59Z alverson $
\newcommand{\mysub}[1]{\vskip 2mm\noindent\textbf{#1:}}
\newcommand{\LQ}{{\rm LQ}}
\newcommand{\MLQ}{M_{\LQ}}
\newcommand{\CL}[1]{$#1\%$~C.L\xdotspace}
\newcommand{\eVdist}{\kern-0.06667em}
\newcommand{\kev}{{\,\text{ke}\eVdist\text{V\/}}}
\newcommand{\mev}{{\,\text{Me}\eVdist\text{V\/}}}
\newcommand{\gevcc}{{\,\text{Ge}\eVdist\text{V\//\ensuremath{c^{2}}}}}
\newcommand{\tev}{{\,\text{Te}\eVdist\text{V\/}}}
\newcommand{\pbi}{\,\text{pb}\ensuremath{^{-1}}}
\newcommand{\fbi}{\,\text{fb}\ensuremath{^{-1}}}
\newcommand {\ptmiss}{\mbox{$\not\hspace{-0.55ex}{P}_t$}}
\newcommand {\misspt}{\mbox{$\not\hspace{-0.55ex}{P}_t$}}
\newcommand{\Cpp}{C\raisebox{2pt}{\scriptsize{++}}}
\newcommand{\sol}{\ensuremath{S_{0}^{L}}}
\newcommand{\sor}{\ensuremath{S_{0}^{R}}}
\newcommand{\tsor}{\ensuremath{\widetilde{S}_{0}^{R}}}
\newcommand{\sil}{\ensuremath{S_{1}^{L}}}
\newcommand{\viiil}{\ensuremath{V_{1/2}^{L}}}
\newcommand{\viiir}{\ensuremath{V_{1/2}^{R}}}
\newcommand{\tviiil}{\ensuremath{\widetilde{V}_{1/2}^{L}}}
\newcommand{\vol}{\ensuremath{V_{0}^{L}}}
\newcommand{\vor}{\ensuremath{V_{0}^{R}}}
\newcommand{\tvor}{\ensuremath{\widetilde{V}_{0}^{R}}}
\newcommand{\vil}{\ensuremath{V_{1}^{L}}}
\newcommand{\siiil}{\ensuremath{S_{1/2}^{L}}}
\newcommand{\siiir}{\ensuremath{S_{1/2}^{R}}}
\newcommand{\tsiiil}{\ensuremath{\widetilde{S}_{1/2}^{L}}}
\newcommand{\be}{\begin{equation}}
\newcommand{\ee}{\end{equation}}
\newcommand{\ud}{\mathrm{d}}
\newcommand{\gimmunnu}{g^{\mu \nu}}
\newcommand{\pimmu}{p^{\mu}}
\newcommand{\pinnu}{p^{\nu}}
\newcommand{\qunnu}{q^{\nu}}
\newcommand{\qummu}{q^{\mu}}
\newcommand{\Tab}[1]{Table~\ref{tab:#1}}
\newcommand{\Tabs}[2]{Tables~\ref{tab:#1} and~\ref{tab:#2}}
\newcommand{\Tabsl}[2]{Tables from ~\ref{tab:#1} to~\ref{tab:#2}}
\newcommand{\Sec}[1]{Sec.~\ref{sec:#1}}
\newcommand{\Secs}[2]{Secs.~\ref{sec:#1} and~\ref{sec:#2}}
\newcommand{\App}[1]{Appendix~\ref{sec:#1}}
\newcommand{\Eq}[1]{Eq.~(\ref{eq:#1})}
\newcommand{\Eqs}[2]{Eqs.~(\ref{eq:#1}) and (\ref{eq:#2})}
\newcommand{\Fig}[2][z]{\mbox{Fig.~\ref{fig:#2}\ifthenelse{\equal{#1}{z}}{}{(#1)}}}
\newcommand{\Figs}[2]{Figs.~\ref{fig:#1} and~\ref{fig:#2}}
\newcommand{\Figsl}[2]{Figs. from~\ref{fig:#1} to~\ref{fig:#2}}
\newcommand{\FIG}[2][z]{\mbox{Figure~\ref{fig:#2}\ifthenelse{\equal{#1}{z}}{}{(#1)}}}
\newcommand{\Chap}[1]{Ch.~\ref{ch:#1}}
\newcommand{\spazio}{\rule{0pt}{15pt}}
\newcommand{\spaziosmall}{\rule{0pt}{4pt}}
\newcommand{\mycaption}[1]{\begin{minipage}{0.9\textwidth}\caption{\small{#1}}\end{minipage}}
\newcommand{\mycaptionshort}[2]{\caption[\small{#1}]{\small{#2}}}
\newcommand{\rnge}{\hbox{$\,\text{--}\,$}}
\newcommand{\DO}{{D{\O}}\xspace}
\newcommand{\cspeed}{\ensuremath{c}}
\newcommand{\alphas}{\ensuremath{\alpha_{S}}}
\newcommand{\lambdaqcd}{\ensuremath{\Lambda_{QCD}}}
\newcommand{\diff}[1]{\ensuremath{\mathrm{d}}#1}
\newcommand{\pthat}{\ensuremath{\hat{p_{\mathrm{T}}}}}
\newcommand{\qcut}{\ensuremath{q_{cut}}}
\newcommand{\PZ}{\ensuremath{{\mathrm{Z}}}\xspace}
\newcommand{\PV}{\ensuremath{{\mathrm{V}}}\xspace}
\newcommand{\photon}{\ensuremath{\gamma}}
\newcommand{\Wp}{\ensuremath{\PW^{+}}\xspace}
\newcommand{\Wm}{\ensuremath{\PW^{-}}\xspace}
\newcommand{\gluon}{\ensuremath{g}}
\newcommand{\zjets}{\ensuremath{\PZ+{\mathrm{jets}}}\xspace}
\newcommand{\zjetsb}{\ensuremath{\PZ+{\mathrm{jets}}}\xspace}
\newcommand{\wjets}{\ensuremath{\PW+{\mathrm{jets}}}\xspace}
\newcommand{\wjetsplus}{\ensuremath{\PW^+\!+{\mathrm{jets}}}\xspace}
\newcommand{\wjetsminus}{\ensuremath{\PW^-\!+{\mathrm{jets}}}\xspace}
\newcommand{\vjets}{\ensuremath{\PV+{\mathrm{jets}}}\xspace}
\newcommand{\vnjets}{\ensuremath{\PV+n~{\mathrm{jets}}}\xspace}
\newcommand{\W}{\PW\xspace}
\newcommand{\MN}{\ensuremath{\mu\nu}}%
\newcommand{\ra}{\ensuremath{\rightarrow}}%
\newcommand{\Wmn}{\ensuremath{\PW \ra \MN}}%
\newcommand{\Zmm}{\ensuremath{\PZ \ra \MM}}%
\newcommand{\Wtn}{\ensuremath{\PW \ra \tau\nu}}%
\newcommand{\Ztt}{\ensuremath{\PZ \ra \tau^+\tau^-}}%
\newcommand{\EN}{\ensuremath{\mathrm{e}\nu}}%
\renewcommand{\EE}{\ensuremath{\mathrm{e}^+\mathrm{e}^-}}%
\newcommand{\Wen}{\ensuremath{\PW \ra \EN}}%
\newcommand{\Zee}{\ensuremath{\PZ \ra \EE}}%

\renewcommand{\ttbar}{\ensuremath{\mathrm{t\bar{t}}}\xspace}
\newcommand{\MT}{\ensuremath{M_{\mathrm{T}}}}

\newcommand{\second}{\text{s}}
\newcommand{\genname}[1]{\texttt{\textbf{#1}}}
\newcommand{\pythia}{\genname{PYTHIA}}
\newcommand{\alpgen}{\genname{AlpGen}}
\newcommand{\sherpa}{\genname{SHERPA}}
\newcommand{\herwig}{\genname{HERWIG}}
\newcommand{\madgraph}{\genname{MADGRAPH}} 
\newcommand{\mcatnlo}{\genname{MC@NLO}}
\newcommand{\amegic}{\genname{AMEGIC++}}
\newcommand{\apacic}{\genname{APACIC++}}
\providecommand{\ROOT} {{\textsc{root}}\xspace}
\providecommand{\MINUIT} {{\textsc{minuit}}\xspace}

\newcommand{\MYANNOT}[1]{%
\fbox{\begin{minipage}{3cm}{\tiny{=$\widehat{\rule{2mm}{0pt}}$.$\widehat{\rule{2mm}{0pt}}$= #1}}\end{minipage}}
}

\newcommand{\pb}{\ensuremath{\mathrm{pb}}}%
\newcommand{\pp}{\ensuremath{\mathrm{pp}}}%
\newcommand{\Wo}{\PW}%
\newcommand{\Zo}{\PZ}%
\newcommand{\rts}{\ensuremath{\sqrt{s}}}%
\renewcommand{\EE}{\ensuremath{\mathrm{e}^+\mathrm{e}^-}}%
\newcommand{\LN}{\ensuremath{\ell\nu}}%
\newcommand{\MW}{\ensuremath{\mathrm{m}_\Wo}}%
\newcommand{\MZ}{\ensuremath{\mathrm{m}_\Zo}}%
\newcommand{\MLL}{\ensuremath{M_{\ell\ell}}}%
\newcommand{\met}{\ensuremath{{E\!\!\!\!/}_{\mathrm{T}}}\xspace}

\newcommand{\Wpmn}{\ensuremath{\Wp \ra \mu^+\nu_\mu}}%
\newcommand{\Wmmn}{\ensuremath{\Wm \ra \mu^-\overline{\nu}_\mu}}%
\newcommand{\ppZmm}{\pp \ra \Zo(\gamma^*) + X \ra \MM + X}%
\newcommand{\ppWmn}{\pp \ra \Wo + X \ra \MN + X}%
\newcommand{\ppWpmn}{\pp \ra \Wp + X \ra \mu^+\nu_\mu + X}%
\newcommand{\ppWmmn}{\pp \ra \Wm + X \ra \mu^-\overline{\nu}_\mu + X}%

\newcommand{\Wpen}{\ensuremath{\Wp \ra \mathrm{e}^+\nu_e}}%
\newcommand{\Wmen}{\ensuremath{\Wm \ra \mathrm{e}^-\overline{\nu}_e}}%
\newcommand{\ppZee}{\pp \ra \Zo(\gamma^*) + X \ra \EE + X}
\newcommand{\ppWen}{\pp \ra \Wo + X \ra \EN + X}%
\newcommand{\ppWpen}{\pp \ra \Wp + X \ra \mathrm{e}^+\nu_e  + X}%
\newcommand{\ppWmen}{\pp \ra \Wm + X \ra \mathrm{e}^-\overline{\nu}_e + X}%

\newcommand{\Wln}{\ensuremath{\Wo \ra \LN}}%
\newcommand{\Wpln}{\ensuremath{\Wp \ra \ell^+\nu}}%
\newcommand{\Wmln}{\ensuremath{\Wm \ra \ell^-\overline{\nu}}}%
\newcommand{\Zll}{\ensuremath{\Zo \ra \ell^+ \ell^-}}%
\newcommand{\ppZll}{\pp \ra \Zo(\gamma^*) + X \ra \ell^+ \ell^- + X}%
\newcommand{\ppWln}{\pp \ra \Wo + X \ra \LN + X}%
\newcommand{\ppWpln}{\pp \ra \Wp + X \ra \ell^+\nu  + X}%
\newcommand{\ppWmln}{\pp \ra \Wm + X \ra \ell^-\overline{\nu} + X}%

\newcommand{\Wptn}{\ensuremath{\Wp \ra \tau^+\nu_\tau}}%
\newcommand{\Wmtn}{\ensuremath{\Wm \ra \tau^-\overline{\nu}_\tau}}%

\newcommand{\Wev}{\Wen}
\newcommand{\Wmv}{\Wmn}
\newcommand{\Ztautau}{\Ztt}
\renewcommand{\MET}{\met}
\newcommand{\gammaZ}{\ensuremath{\gamma^{*}\Zo}}
\newcommand{\gammaZmm}{\mbox{$ \gamma^{*}/\Zo\rightarrow \MM$}}
\newcommand{\gammaZee}{\mbox{$ \gamma^{*}\Zo\rightarrow e^{+}  e^{-}$}}
\newcommand{\gammaZtt}{\mbox{$ \gamma^{*}\Zo\rightarrow \tau^{+}  \tau^{-}$}}
\newcommand{\gammaZll}{\mbox{$ \gamma^{*}\Zo\rightarrow \ell^{+}  \ell^{-}$}}
\newcommand{\invpb}{\mbox{$\textrm{pb}^{-1}$}}
\newcommand{\invnb}{\mbox{$\textrm{nb}^{-1}$}}
\newcommand{\mz}{\mbox{$ m_{Z}$}}
\newcommand{\hta}{\mbox{$ \eta$}}
\newcommand{\fh}{\mbox{$ \phi$}}
\newcommand{\etot}{\mbox{$ \epsilon_{tot}$}}
\newcommand{\eclustering}{\mbox{$ \epsilon_{clustering}$}}
\newcommand{\etracking}{\mbox{$ \epsilon_{tracking}$}}
\newcommand{\egsfele}{\mbox{$ \epsilon_{gsfele}$}}
\newcommand{\epreselection}{\mbox{$ \epsilon_{preselection}$}}
\newcommand{\eisolation}{\mbox{$ \epsilon_{isolation}$}}
\newcommand{\eclassification}{\mbox{$ \epsilon_{classification}$}}
\newcommand{\eelID}{\mbox{$ \epsilon_{elID}$}}
\newcommand{\etrigger}{\mbox{$ \epsilon_{trigger}$}}

\newcommand{\DE}{$\Delta\eta_{in}$}
\newcommand{\DP}{$\Delta\phi_{in}$}
\newcommand{\SEE}{$\sigma_{\eta\eta}$~}
\newcommand{\SEP}{$\sigma_{\eta\phi}$}
\newcommand{\SPP}{$\sigma_{\phi\phi}$}
\newcommand{\SXY}{$\sigma_{XY}$}
\newcommand{\pth}{\hat{p}_{\perp}}
\newcommand{\Pt}{\ensuremath{p_{T}}~}
\newcommand{\Et}{\ensuremath{E_{T}}~}
\newcommand{\Lint}{\ensuremath{{\cal L}_{\mathrm{int}}}}

\newcommand{\IRel}{I_{\mathrm{iso}}}
\mathchardef\mhyphen="2D
\newcommand{\NBJET}{n_{\mathrm{jet}}^{\mathrm{b\mhyphen tagged}}}

\newcommand{\RW}{\ensuremath{R^\pm_\PW}}
\newcommand{\AW}{\ensuremath{A_\PW}}
\cmsNoteHeader{EWK-10-012} 
\title{Jet Production Rates in Association with W and Z Bosons\\
in pp Collisions at $\sqrt{s} = 7$~TeV}

\date{\today}

\abstract{
Measurements of jet production rates in association with W and Z bosons for jet
transverse momenta above 30~\GeV are reported, using a sample of proton-proton
collision events recorded by CMS at $\sqrt{s} = 7$~TeV, corresponding
to an integrated luminosity of $36 \pbi$.
The study includes the measurement of the normalized inclusive rates
of jets $\sigma(\PV+\ge n~{\mathrm{jets}})/\sigma(\PV)$, where \PV\ represents either a \W or a \Z.
In addition, the ratio of W to Z
cross sections and the W charge asymmetry as a function of the number of associated jets are measured.
A test of
Berends--Giele scaling at $\sqrt{s} = 7$~TeV is also presented. The measurements provide
a stringent test of perturbative-QCD calculations and are sensitive
to the possible presence of new physics. The results are in
agreement with the predictions of a simulation
that uses explicit matrix element calculations for final
states with jets.
}
\hypersetup{%
pdfauthor={CMS Collaboration},%
pdftitle={Jet Production Rates in Association with W and Z Bosons in pp Collisions at sqrt(s) = 7 TeV},%
pdfsubject={CMS},%
pdfkeywords={CMS, physics}}

\maketitle 

\graphicspath{{Figures/}}

\newlength{\sizeFigThree}
\setlength{\sizeFigThree}{0.31\textwidth}
\newlength{\sizeFigTwo}
\setlength{\sizeFigTwo}{0.48\textwidth}

\section{Introduction}
\par
The study of jet production in association with a \W or \Z vector boson
(denoted by '\PV' in this paper) provides a stringent test of perturbative QCD
calculations. The presence of a vector boson provides a clear
signature of the process and allows comparison of different scattering
amplitudes with respect to inclusive multijet production, which is
dominated by gluon scattering. In addition, a precise measurement of
the W (Z) + $n$ jets cross section is essential since the production
of vector bosons with jets constitutes a background in searches for
new physics and for studies of the top quark.
At present, next-to-leading-order (NLO) predictions are available for
\vnjets, with $n$ up to four~\cite{1108.2229v2,PhysRevLett.106.092001,
  PhysRevD.80.074036, PhysRevD.80.094002, PhysRevD.82.074002}.
\par
The \Wp to \Wm cross-section ratio for an associated jet multiplicity
is also predicted by the theory~\cite{kom} and larger than unity.
Deviations of this ratio from the expected value may point to new
processes that produce W bosons in the final state.
\par
The CDF, \DZERO and ATLAS Collaborations have reported measurements of
W (Z) + $n$ jets production in~\cite{CDFW,CDFZ,D0W,D0Z,ATLASW}.
\par
This paper presents results obtained with the 2010 data sample of
the Compact Muon Solenoid (CMS) experiment at the Large Hadron Collider (LHC), based
on an integrated luminosity of $35.9\pm 1.4~\pbi$~\cite{LUMI}, collected in
proton-proton collisions at 7 TeV.
To reduce systematic uncertainties associated with the integrated luminosity
measurement, the jet energy scale (JES), the lepton reconstruction,
and the trigger efficiencies, we measure the \vnjets cross sections relative to
the inclusive \W and \Z cross sections, $\sigma(\PV+\ge n~{\mathrm{jets}})/\sigma(\PV)$.
A clear advantage of using ratios of cross sections is also that
theoretical uncertainties tend to cancel, improving the
robustness of the results.
We also measure the ratio of the W to Z
cross sections and the W charge asymmetry as a function of
the associated jet multiplicity.
Finally, we test the Berends--Giele scaling hypothesis~\cite{BGscaling}.
To compare the measurements with the predictions of the theoretical
calculations, the results are presented at ``particle level'',
unfolding the detector efficiency and resolution.
\par

\section{The CMS Experiment}
\label{sec:cms}
The central feature of the CMS apparatus is a superconducting solenoid of 6~m internal diameter,
providing an axial magnetic field of 3.8~T. Within the field volume are a silicon pixel and
strip tracker, an electromagnetic calorimeter (ECAL), and a brass/scintillator hadron calorimeter
(HCAL). Muons are detected in gas-ionization detectors embedded in the steel return yoke.
\par
A right-handed coordinate system is used in CMS, with the origin at the
nominal interaction point, the $x$-axis pointing to the centre of
the LHC ring, the $y$-axis pointing up (perpendicular to the LHC plane),
and the $z$-axis along the anticlockwise-beam direction. The polar
angle $\theta$ is measured from the positive $z$-axis and the
azimuthal angle $\phi$ is measured from the positive $x$-axis in the
$x$-$y$ plane.
The pseudorapidity is defined as $\eta = -\ln [\tan(\theta/2)]$.
\par
The inner tracker contains 1440 silicon pixel and 15148 silicon 
strip detector modules. It measures charged particle trajectories in
the pseudorapidity range $|\eta| \le$~2.5. 
The ECAL consists of nearly 76000 lead
tungstate crystals that provide coverage for $|\eta| \le$~1.479 in a cylindrical barrel region (EB) and 
1.479~$\le |\eta|\le$~3.0 in the two endcap regions (EE). Preshower
detectors, each consisting of two planes of silicon sensors
interleaved with a total of 3 radiation length of lead, are located in front of both EEs. 
The HCAL is a sampling device with brass as the passive material
and scintillator as the active material. The combined calorimeter cells are grouped in projective
towers of granularity $\Delta \eta \times \Delta \phi = 0.087 \times 0.087$ at central rapidities and
$ \Delta \eta \times \Delta \phi \approx 0.17\times 0.17$ at forward rapidities.
Muons are detected in the pseudorapidity range $|\eta| \le$~2.4, with detection planes based on
three technologies: drift tubes, cathode strip chambers, and resistive plate chambers. 
In addition to the barrel and endcap detectors, CMS has extensive forward calorimetry.
The first level (L1) of the trigger system, composed of custom hardware processors, is
designed to select event candidates in less than $3.2$ $\mu$s using information from the
calorimeters and muon detectors. The high level trigger (HLT) processor farm further reduces
the event rate to a few hundred hertz, before data storage.
A more detailed description of CMS can be found elsewhere~\cite{:2008zzk}.

\section{Data and Simulation Samples}
\label{sec:datamc}
The L1 trigger system selected electrons with an 
energy deposit in the ECAL of at least 5~GeV or 8~GeV, depending on the
luminosity conditions, and muons with a transverse 
momentum exceeding 7~GeV. The events were then filtered by the HLT with
algorithms that evolved in response to the rapid rise of 
the LHC luminosity during 2010. The \PT threshold of the electrons and
muons was adjusted periodically, to cope with the increasing instantaneous luminosity. 
The largest sample of electrons was collected
with an online requirement $\PT \ge 17$~\GeV. 
For muons, most data were collected with a threshold of
$\PT \ge 15$~\GeV.

Simulated Monte Carlo (MC) samples are used
for data/simulation comparison and to unfold the jet
multiplicity distributions.
Backgrounds are estimated from data, as explained below.

Samples of simulated events with a \W or a \Z boson are generated with the
\MADGRAPH~{4.4.13} \cite{Maltoni:2002qb} event generator, interfaced to the  
\PYTHIA~{6.422}~\cite{Sjostrand:2006za} program for parton shower
simulation. The set of parton distribution functions used is
\textsc{CTEQ6L1}~\cite{CTEQ}. The \MADGRAPH generator
produces parton-level events with a vector boson and up to
four partons on the basis of a matrix-element calculation.
This sample serves as the baseline for comparisons with data.
Additional samples of W and Z events are generated with \PYTHIA.  
For jet multiplicities greater than
one, \MADGRAPH is expected to be more accurate since it uses the exact
matrix-element calculation, while in
\PYTHIA\ only the hardest emission reproduces the exact matrix-element calculation.
Top-pair (\ttbar) and single-top production processes
are generated with \MADGRAPH. Processes of multijet, $\gamma$+jets,
b-hadron and c-hadron decays to final states with electrons and muons are generated with
\PYTHIA.

The full list of simulated samples is given in 
Table~\ref{tab:mcsamples}. In order to compare with the data distributions, 
the simulation samples are normalized to the cross sections times
the integrated luminosity, using or NLO or next-to-next-to-leading-order
(NNLO) cross sections \cite{Melnikov:2006kv,Campbell:2009ss,Kleiss:1988xr}, or the
leading-order (LO) cross sections from the MC generator, as
reported in the table. 

Generated events are processed through a full detector simulation
program based on GEANT4 \cite{Agostinelli:2002hh,Allison:2006ve}, followed
by a detailed emulation of the trigger and event reconstruction.
Minimum-bias events are superimposed on the generated events to reproduce the
distribution of multiple proton-proton collisions in the same
bunch crossing (pileup) observed in 2010. 
A signal sample without pileup is used
for purpose of comparison.   The PYTHIA parameters for the underlying
event are set to the Z2 tune~\cite{UEPAS}, which is a modification
of the Z1 tune described in Ref.~\cite{Field:2010bc}.
Comparisons are also made to the D6T tune~\cite{Skands:2010ak}.

\begin{table}[htb]
\caption{Summary of simulated datasets for the various signal and
  background processes used in this analysis. The requirements applied
  to leptons $\PT$ and $\eta$, dilepton invariant mass ($\MLL$), and
  transverse momentum of the hard interaction ($\hat{\PT}$), are shown,
  and the corresponding cross section is given. \label{tab:mcsamples} }
\begin{center}
{\footnotesize
\begin{tabular}{| l | c | c | c | c|}
\hline
 Process &  Generator & Kinematic selection & $\sigma$~(pb)\\
\hline\hline
$\PW \ra \ell\nu $                &  \MADGRAPH & no selection & 3.1\ten{4} (NNLO) \\ 
$\PZ \ra \ell^+\ell^- $           &  \MADGRAPH & $\MLL>50$ GeV & 3.0\ten{3} (NNLO) \\ 
$\ttbar$                      &  \MADGRAPH & no selection & 1.6\ten{2}
(NLO) \\
Single-top $\mathrm{tW}$ channel         &  \MADGRAPH & no selection &  1.1\ten{1}  (LO)  \\
Single-top s and t channels &  \MADGRAPH & no selection &  3.5  (NLO)  \\
$\Wen$                                        & \PYTHIA &  $|\eta_e|<2.7$  &  8.2\ten{3} (NNLO)    \\
$\Wmn$                                        & \PYTHIA &  $|\eta_\mu|<2.5$ & 7.7\ten{3} (NNLO)   \\
$\Wtn$                                        & \PYTHIA &  no selection & 1.0\ten{4} (NNLO)  \\
$\PZ \ra \ell^+\ell^- $                         & \PYTHIA & $\MLL>20$ GeV & 5.0\ten{3} (NNLO) \\
Inclusive $\mu$ QCD multijet                          & \PYTHIA &
$\hat{\PT}>20$ GeV, $\PT^\mu>10$ GeV, $|\eta_{\mu}|<2.5$ & 3.4\ten{5} (LO)  \\
EM-enriched QCD  multijet                             & \PYTHIA &  20 GeV $< \hat{\PT}< $ 170 GeV & 5.4\ten{6}  (LO) \\
$\mathrm{b}/\mathrm{c}\rightarrow \mathrm{e}$ & \PYTHIA &  20 GeV $< \hat{\PT}<$ 170 GeV & 2.6\ten{5} (LO) \\
$\gamma$+jet                                  & \PYTHIA &  no selection                & 8.5\ten{7} (LO) \\
\hline
 \end{tabular}
}
 \end{center}
\end{table}

\section{Signal Selection}
\label{sec:selection}
Signal selection begins with the identification of a charged
lepton, either an electron or a muon,  
with $\PT > 20$~\GeV.
This lepton, which will be called the ``leading lepton'', must
geometrically match the object that triggered the event readout.

For electron candidates, we require that the ECAL cluster lies in
the fiducial region $|\eta| < 2.5$, with the exclusion of the region
$1.4442 < |\eta| < 1.566$. This exclusion allows us to reject
electrons close to the 
barrel/endcap transition and electrons in the first endcap 
trigger tower, which lies in the shadow of cables and services.
A series of quality requirements, following the standard established
by the measurement of the inclusive \W and \Z cross 
sections~\cite{Khachatryan:2010xn}, are then applied to the electron, as 
briefly described below.
\par
For each electron candidate, a supercluster in ECAL is defined in
order to correct for the potential underestimation of the energy due
to bremsstrahlung.  Thus the electron cluster is combined with the
group of single clusters attributed to bremsstrahlung photons
generated in the material of the tracker. 
Electrons are first selected based on
the spatial matching between the ECAL supercluster and the silicon
detector track in the $\eta$ and $\phi$ coordinates, on the
supercluster energy distribution in the $\eta$ 
direction, and on the energy leakage  into the HCAL detector. 
To reduce the contamination from converted photons, a minimal track transverse
impact parameter significance is required.
Electrons are rejected if no associated track hits are found in the first
tracker layers or if a conversion partner candidate is
found.
\par
To reduce further the contamination from misidentified electrons and 
hadronic decays, we select electrons isolated from hadronic activity. 
This selection is based on maximum values allowed for three isolation variables. 
The variables are computed relative to the electron \Et and consist of
the sums of track $\PT$ in the tracker, energy deposits in
ECAL, and energy deposits in HCAL, respectively.
The sums are computed inside the cone $\Delta R = \sqrt{(\Delta
\eta)^{2} + (\Delta \phi)^{2}} < 0.3$,  where $\Delta \eta$ and $\Delta
\phi$ are the differences in the pseudorapidity and azimuthal angle
between the electron and the track or energy deposit, with
an inner exclusion region that removes the electron contribution.
\par
For the leading electron, the values of the different quality requirements are chosen such that they
correspond to a lepton efficiency of about 80\%, as evaluated with 
the \MADGRAPH+\PYTHIA simulated sample described in Section~\ref{sec:efficiency}. 
\par
After identifying the leading electron, we search for a second electron 
candidate called the ``second leading electron'', within the ECAL fiducial volume and with \pt$> 10$~GeV. The quality requirements for the
second leading electron are tuned to provide an electron efficiency of
about 95\%, as evaluated with the \MADGRAPH+\PYTHIA simulated sample. If such a
second leading electron is found, and its invariant 
mass with the first leading electron $\MLL$ lies between $60~\GeV$ and 
$120~\GeV$,  the event is assigned to the \zjetsb~sample. If such a second leading electron is not found, 
the event is assigned to the \wjets sample, thereby ensuring 
that there is no overlap between the two samples. Events including a muon
with \pt $ > 15~\GeV$ and $|\eta| < 2.4$ are rejected from the
\wjets electron sample to reduce \ttbar contamination.
\par
The muon reconstruction and identification are identical to 
that used for the measurement of the \W and \Z
cross sections~\cite{Khachatryan:2010xn}.
A relative isolation variable,
$I = \sum ( \pt^{\mathrm{track}} + \ET^{\mathrm{HCAL}} +
\ET^{\mathrm{ECAL}} ) / \pt^{\mu} $ is defined,
which includes the $\PT$ for tracks, ECAL, and HCAL towers
in a cone $\Delta R < 0.3$ around the muon 
direction.
The muon and its energy deposits are excluded from the sum.
A muon is considered to be isolated if $I < 0.15$.
The \vjets muon event selection starts by requiring the presence of an 
isolated muon in the region  $|\eta| < 2.1$ with \pt~$> 20$~\GeV.  It must be a
high-quality muon as described in~\cite{Khachatryan:2010xn} with an impact parameter 
in the transverse plane $|d_{xy}| < 2$~mm,
to suppress cosmic-ray background. 
As for the electron channel,  we then search for a 
second leading muon with \pt$ > 10$~\GeV and
 $|\eta| < 2.4$, such that the dimuon invariant mass lies between 
 60\GeV to 120\GeV. If such a second leading muon is (is not) found,
 the event is assigned to the \zjetsb~(\wjets) sample. An electron
 veto is not applied in the selection of W decays into muons as it
 would significantly lower the efficiency and increase the systematic
 uncertainty, because of fake electrons.
\par
For the \wjets samples in both decay modes, we compute the missing transverse energy \MET
using a particle-flow (PF) algorithm~\cite{ref:pf} that reconstructs
individually each particle in the event based on information from all
relevant subdetectors. We then use it to calculate the transverse mass, 
$\MT = \sqrt{2\PT\MET (1-\cos\Delta\phi)}$, 
where $\Delta\phi$  is the angle in the $x$-$y$ plane
between the directions of the lepton \PT and the $\MET$, and select
events with \MT~$> 20$~\GeV.
\par 
For the charge asymmetry measurement, the charge assignment of the lepton is used 
to select  \wjetsplus and \wjetsminus candidates. For electrons, three
algorithms are used to determine their charge~\cite{Wasym}: 
the curvature of the electron track reconstructed by a
Gaussian-sum-filter algorithm~\cite{GSF}, 
the curvature of the silicon detector track associated with the electron, and 
the difference in $\phi$ between the track direction at
the interaction vertex and the supercluster measured by the ECAL. In order to
reduce uncertainties related to charge misassignment, we 
reject the event if the charge value differs amongst the three methods.

\section{Jet Rates}
\label{subsec:jetreco}
\par
Jets are reconstructed from the particle collection created
with the particle-flow  algorithm  and are
formed with the anti-$k_T$ clustering algorithm~\cite{ref:antikt} 
with a size parameter of $R = 0.5$.   Jet energy corrections (JEC)
are applied  to account for the jet energy response as a function 
of~$\eta$ and \ET~\cite{jme-10-010}.
\par
We require the jet to have $|\eta| < 2.4$ to be in the
tracker acceptance, and \Et $>30$\GeV.  Jets are required to satisfy
identification criteria that eliminate jets originating from or  seeded by  
noisy channels in the hadron calorimeter~\cite{Chatrchyan:2009hy}.

The pileup and the underlying event affect the jet 
counting by contributing additional energy to the measured jet energy and
therefore ``promoting'' jets  above the \ET threshold for jet counting.
To minimize the uncertainty due to pileup and the underlying event,
the \ET threshold is set at 30~GeV. 
The average number of pileup events in the data sample is 2.7.
Their effect is taken into account by evaluating  event-by-event the
energy not related to the hard-interaction activity
~\cite{fastjet1,fastjet2}.  
This amount is subtracted from each jet~\cite{jme-10-010}. 
Removing the average energy due to pileup does not remove the jets from
pileup interactions.
In a simulated sample, this additional contribution to the jet count, however, is found to be
negligible for the chosen jet \ET\ threshold. 

Electrons can be reconstructed as jets or can overlap
with a jet.  Therefore, jets that fall within $\Delta R < 0.3$
of an electron from \W or \Z decay are not included
in the jet count. Muons can also overlap with a jet, thus, each selected
muon is matched to one of the particles reconstructed by the PF
algorithm and excluded from jet clustering. 

One of the most important backgrounds in the \W sample at high jet
multiplicity comes from \ttbar events. These contain two b-quark jets.
We count the number of b-tagged jets, $\NBJET$,
with a  tagging algorithm that requires at least two tracks in the
jet with a significance on the transverse impact parameter greater
than 3.3. The algorithm parameters chosen correspond
to a working point with an efficiency of about 62\% and a mistag rate
of about 2.9\%. The efficiency is measured from data using a sample of
\ttbar events with fully leptonic final states. The mistag rate is
averaged over light jets in simulated W and top events, and corrected
for measured differences between data and simulation~\cite{btv-11-001}.  
The $\NBJET$ value is then used in the fitting
method to extract the number of signal events, as described in Section~\ref{sec:fit}.

\begin{figure}
  \centering
\includegraphics[width=\sizeFigTwo]{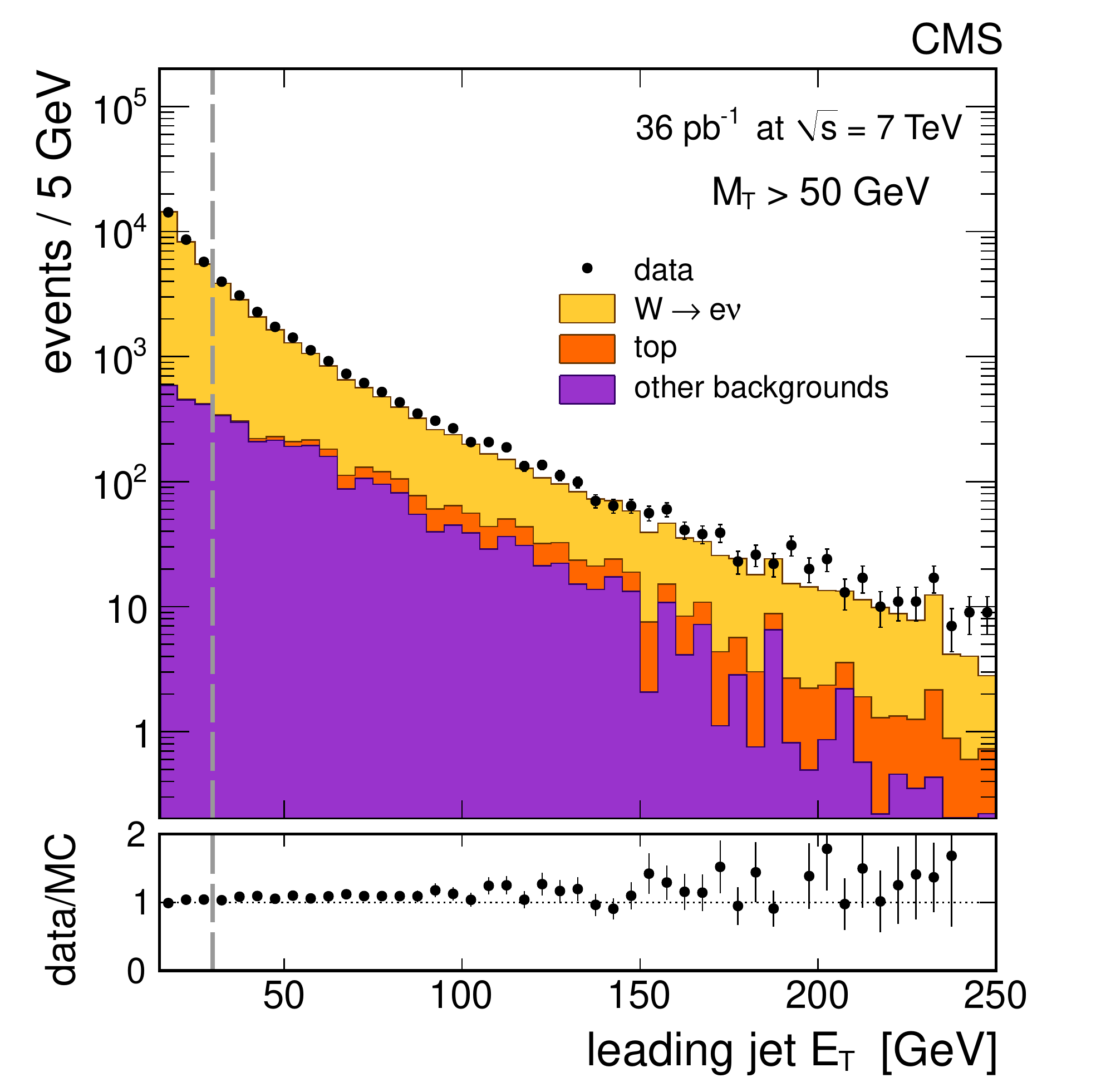}
\includegraphics[width=\sizeFigTwo]{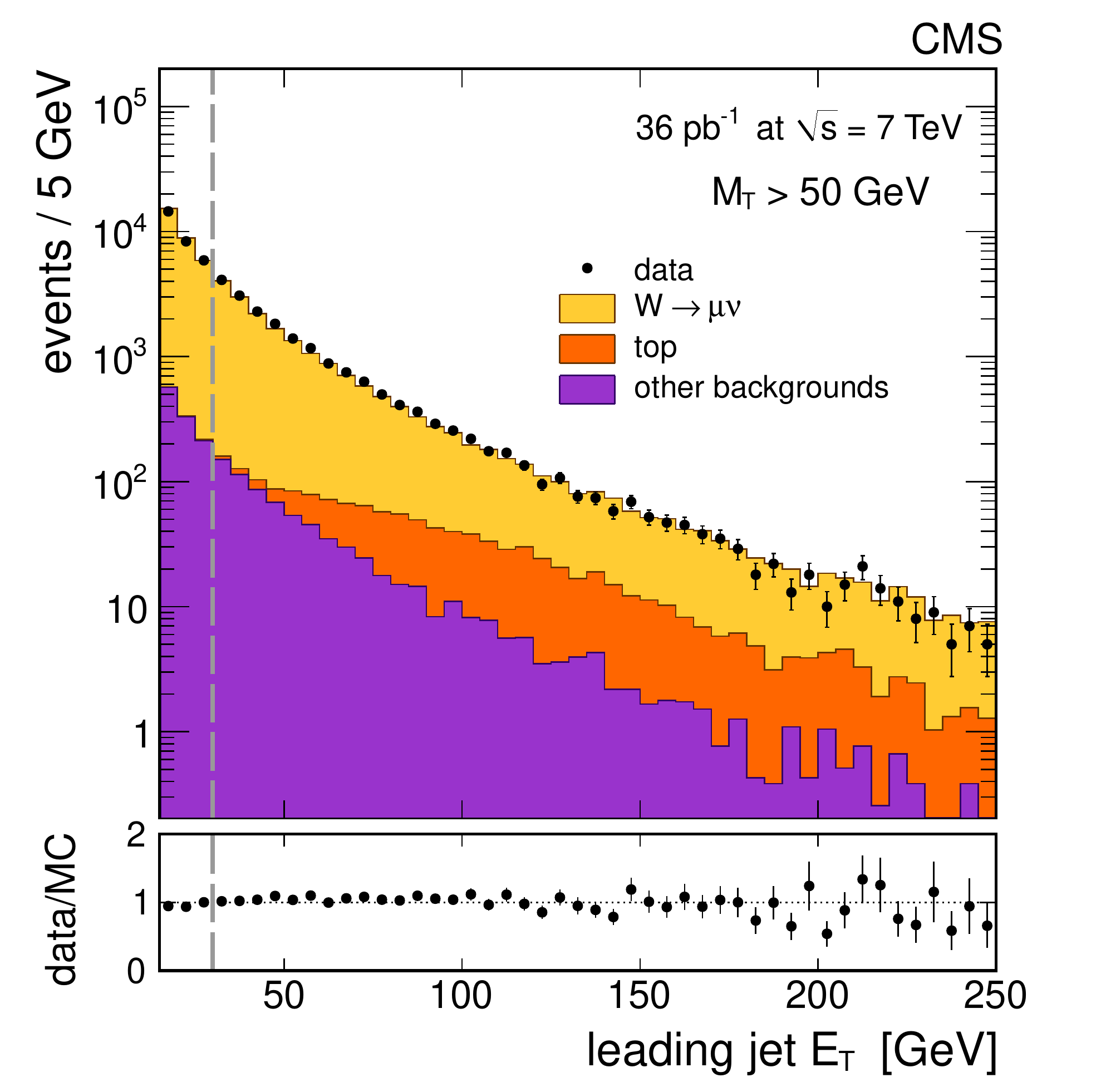}
  \caption{
Distributions of the  \ET for the leading jet
in the $W + \ge 1$~jet sample for the electron channel (left)
and for the muon channel (right), before the requirement
of $\ET > 30$~\GeV (shown by the vertical dotted line) is imposed for 
counting jets. Points with error bars are data,
histograms represent simulation of the signal (yellow), \ttbar and single
top backgrounds (orange) and other backgrounds (purple). The last 
includes multijet events, \Z events, and events in which the \W decays to a
channel other than the signal channel. Error bars represent the
statistical uncertainty on the data. 
The ratio between the data and the simulation is shown in the lower plots.
\label{fig:leadjetPt_W}}
  \centering
\includegraphics[width=\sizeFigTwo]{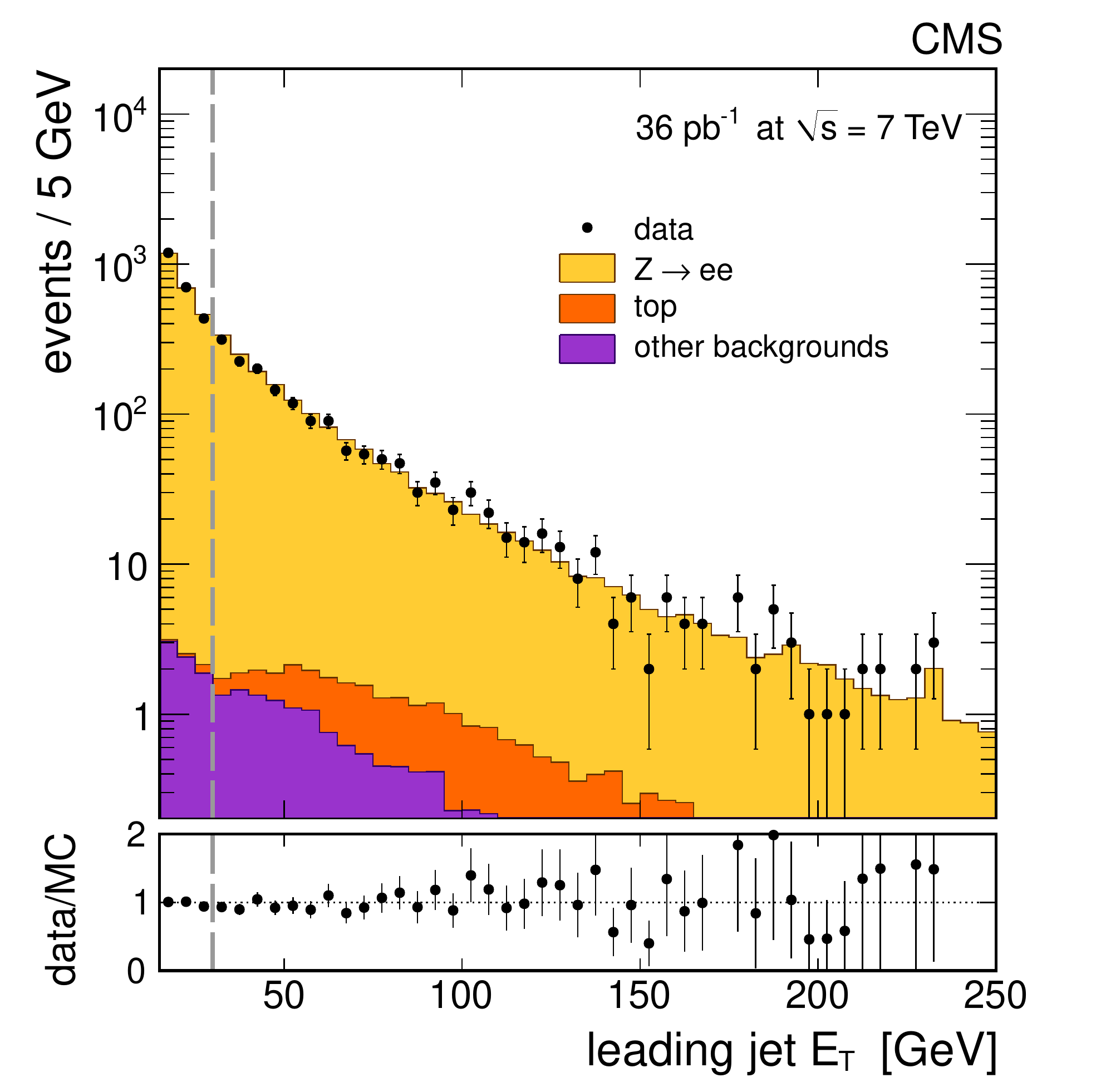}
\includegraphics[width=\sizeFigTwo]{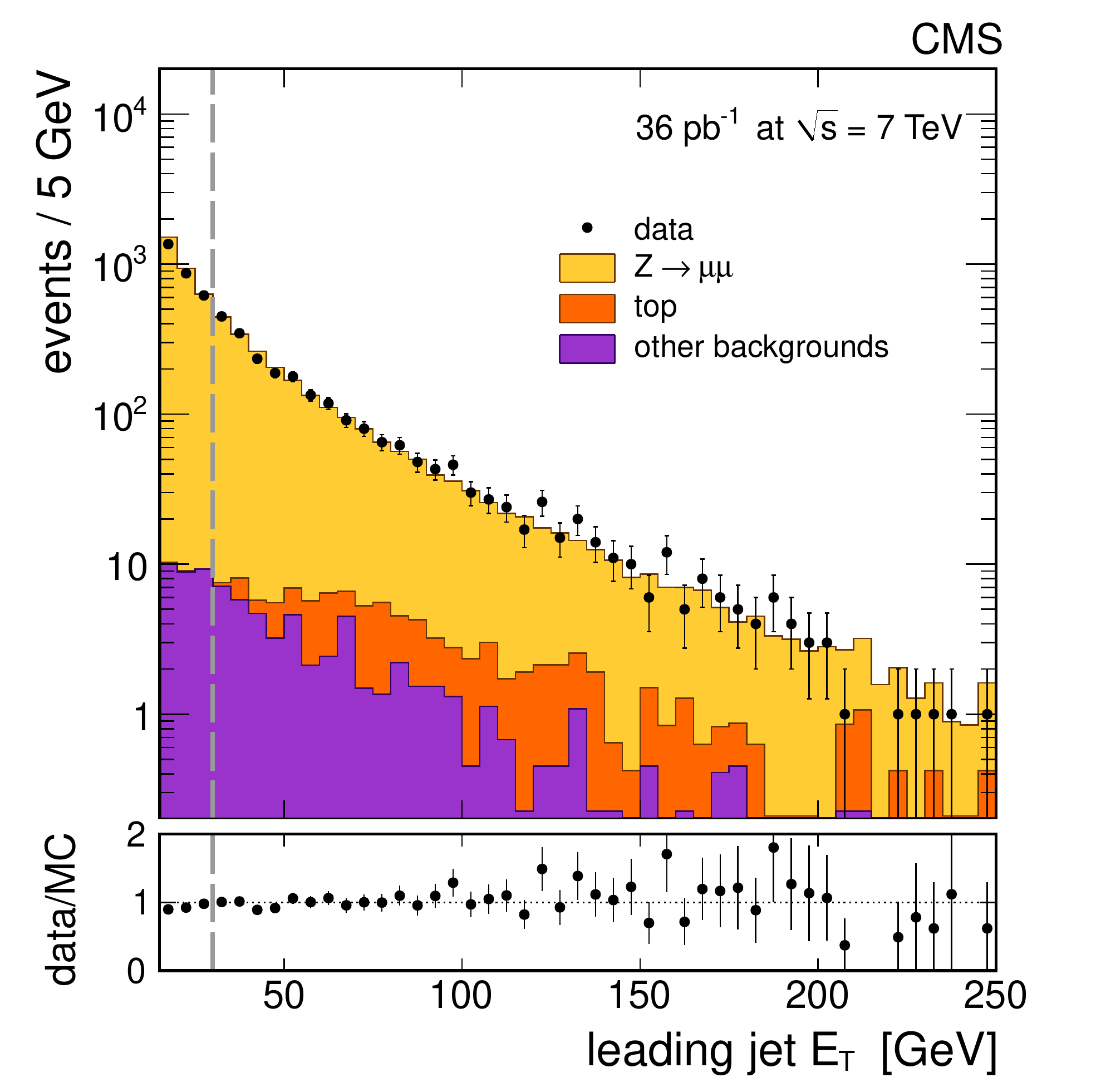}
  \caption{
Distributions of the  \ET for the leading jet
in the $Z + \ge 1$~jet sample for the electron channel (left) and for the muon channel (right), before the requirement
of $\ET > 30$~\GeV (shown by the vertical dotted line) is imposed for 
counting jets. 
Points with error bars are data, histograms represent simulation of
the signal (yellow), \ttbar and single top 
backgrounds (orange) and other backgrounds (purple). The last includes 
multijet events, \W events, and events in which the \Z decays to a
channel other than the signal channel. Error bars represent the statistical
uncertainty on the data. 
The ratio between the data and the simulation is shown in the lower plots.
\label{fig:leadjetPt_Z}}
\end{figure}

\begin{figure}
  \centering
\includegraphics[width=\sizeFigTwo]{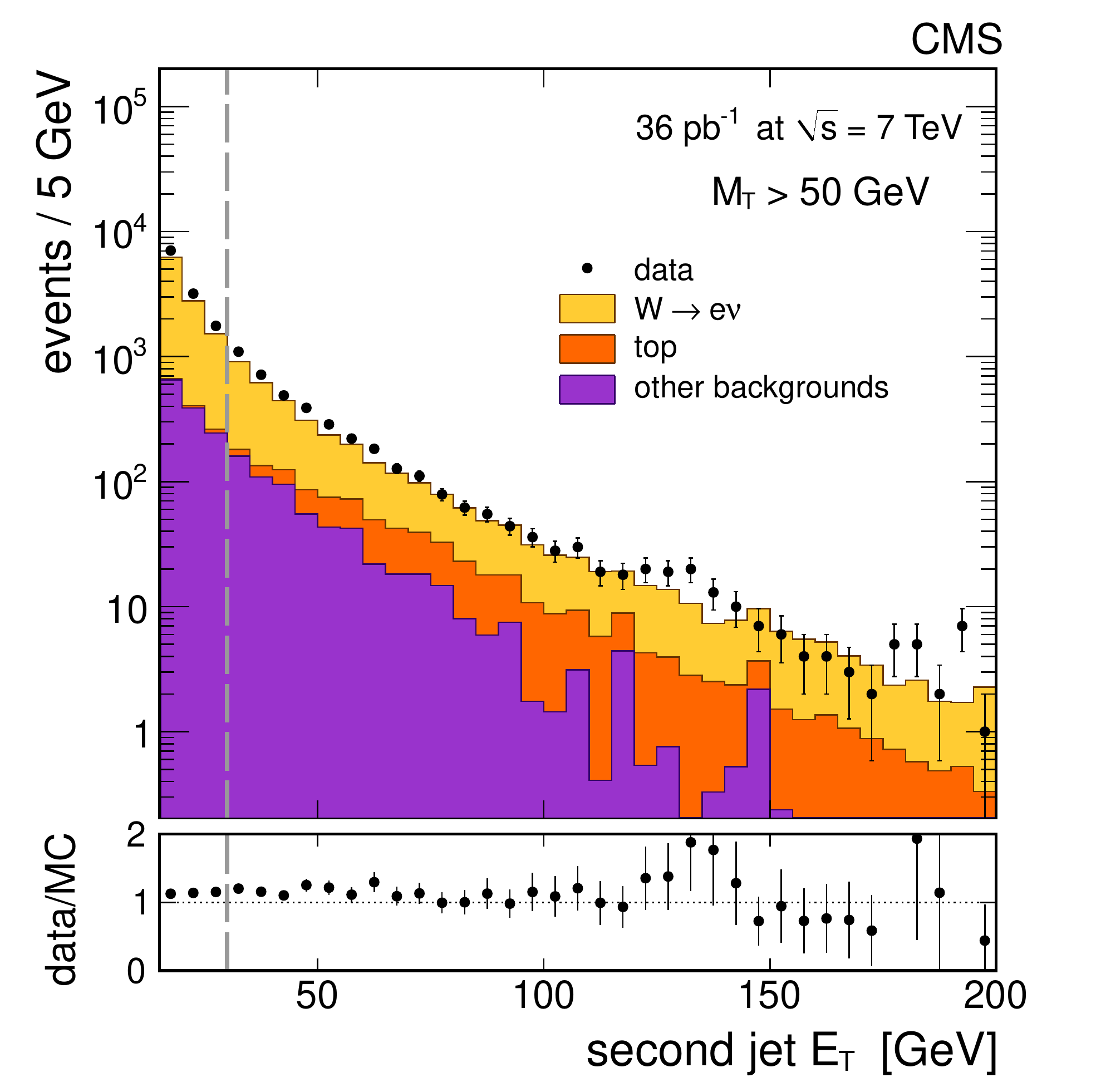}
\includegraphics[width=\sizeFigTwo]{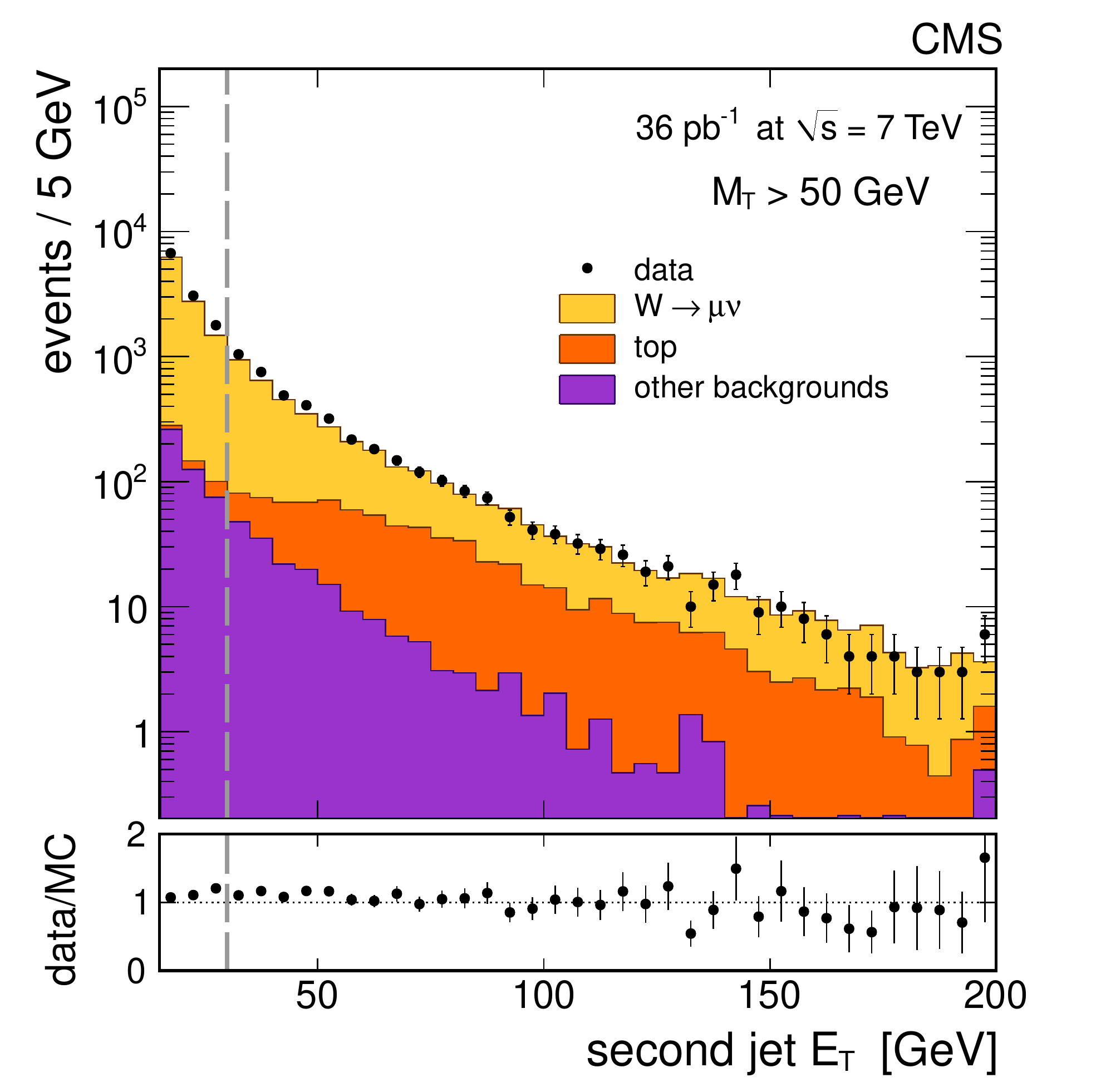}
  \caption{
Distributions of the  \ET for the second leading jet
in the $\PW + \ge 2$~jets sample for the electron channel (left)
and for the muon channel (right), before the requirement
of $\ET > 30$~\GeV (shown by the vertical dotted line) is imposed for 
counting jets. Points with error bars are data,
histograms represent simulation of the signal (yellow), \ttbar and single top
backgrounds (orange) and other backgrounds (purple). The last
includes multijet events, \Z events, and events in which the \W decays to a
channel other than the signal channel. Error bars represent the
statistical uncertainty on the data.
The ratio between the data and the simulation is shown in the lower plots.
\label{fig:secondjetPt_W}}
  \centering
\includegraphics[width=\sizeFigTwo]{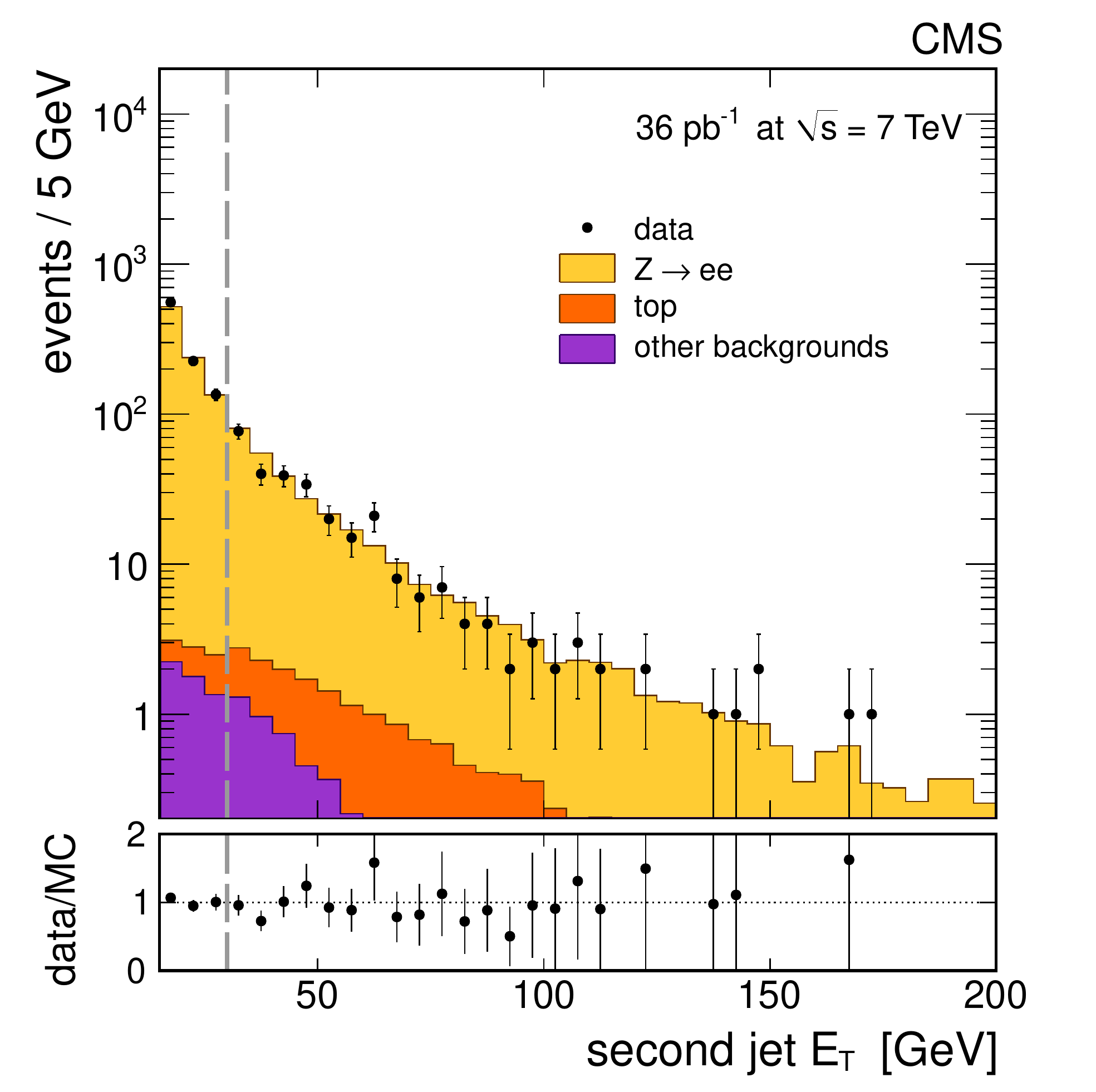}
\includegraphics[width=\sizeFigTwo]{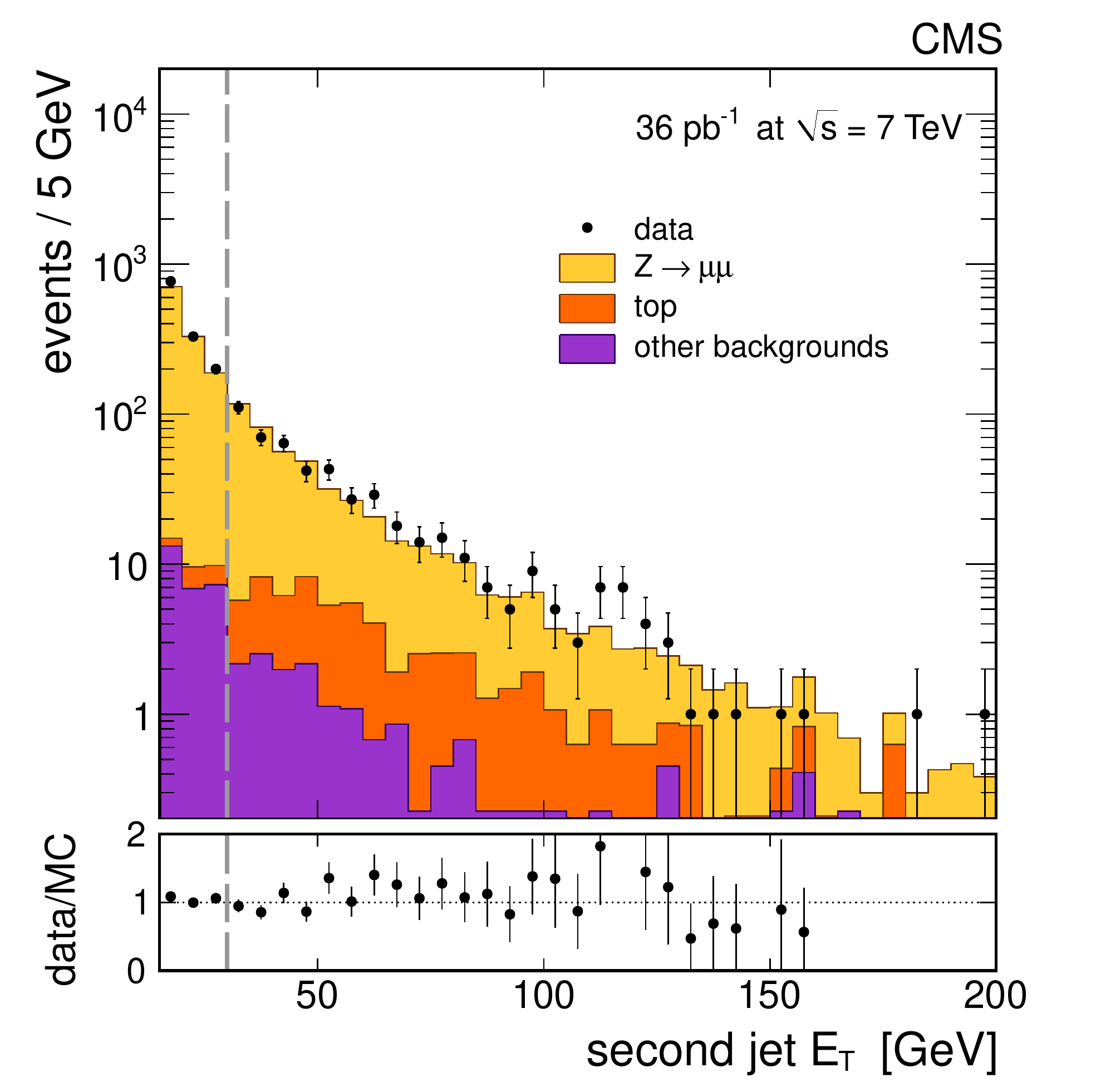}
  \caption{
Distributions of the \ET for the second leading jet
in the $\PZ + \ge 2$~jets sample for the electron channel (left) and for the muon channel (right), before the requirement
of $\ET > 30$~\GeV (shown by the vertical dotted line) is imposed for 
counting jets. 
Points with error bars are data, histograms represent simulation of
the signal (yellow), \ttbar and single
top backgrounds (orange) and other backgrounds (purple). The last includes 
multijet events, \W events, and events in which the \Z decays to a
channel other than the signal channel. Error bars represent the statistical
uncertainty on the data.
The ratio between the data and the simulation is shown in the lower plots.
\label{fig:secondjetPt_Z}}
\end{figure}

The observed  transverse energy distributions of the leading jet
in the $\PW + \ge 1$~jet and $\PZ + \ge 1$~jet samples are shown in Figs.~\ref{fig:leadjetPt_W} and~\ref{fig:leadjetPt_Z},
respectively.
Figures~\ref{fig:secondjetPt_W} and~\ref{fig:secondjetPt_Z} show the same
distributions for the second-leading jet in the $W + \ge 2$~jets and $Z
+ \ge 2$~jets samples. 
In addition to the selection described in Section~\ref{sec:selection}, for the
transverse energy distributions, \W boson candidate are
required to have $\MT > 50$~\GeV, thus reducing significantly the QCD
multijet contamination. The $\ttbar$ background is larger in the muon
than in the electron channel because an electron veto is not
applied, as explained in Section~\ref{sec:selection}.
The data are in good agreement with the \MADGRAPH + \PYTHIA parton shower predictions 
normalized to the NNLO cross sections. 

\par
\begin{figure}
  \centering
\includegraphics[width=\sizeFigTwo]{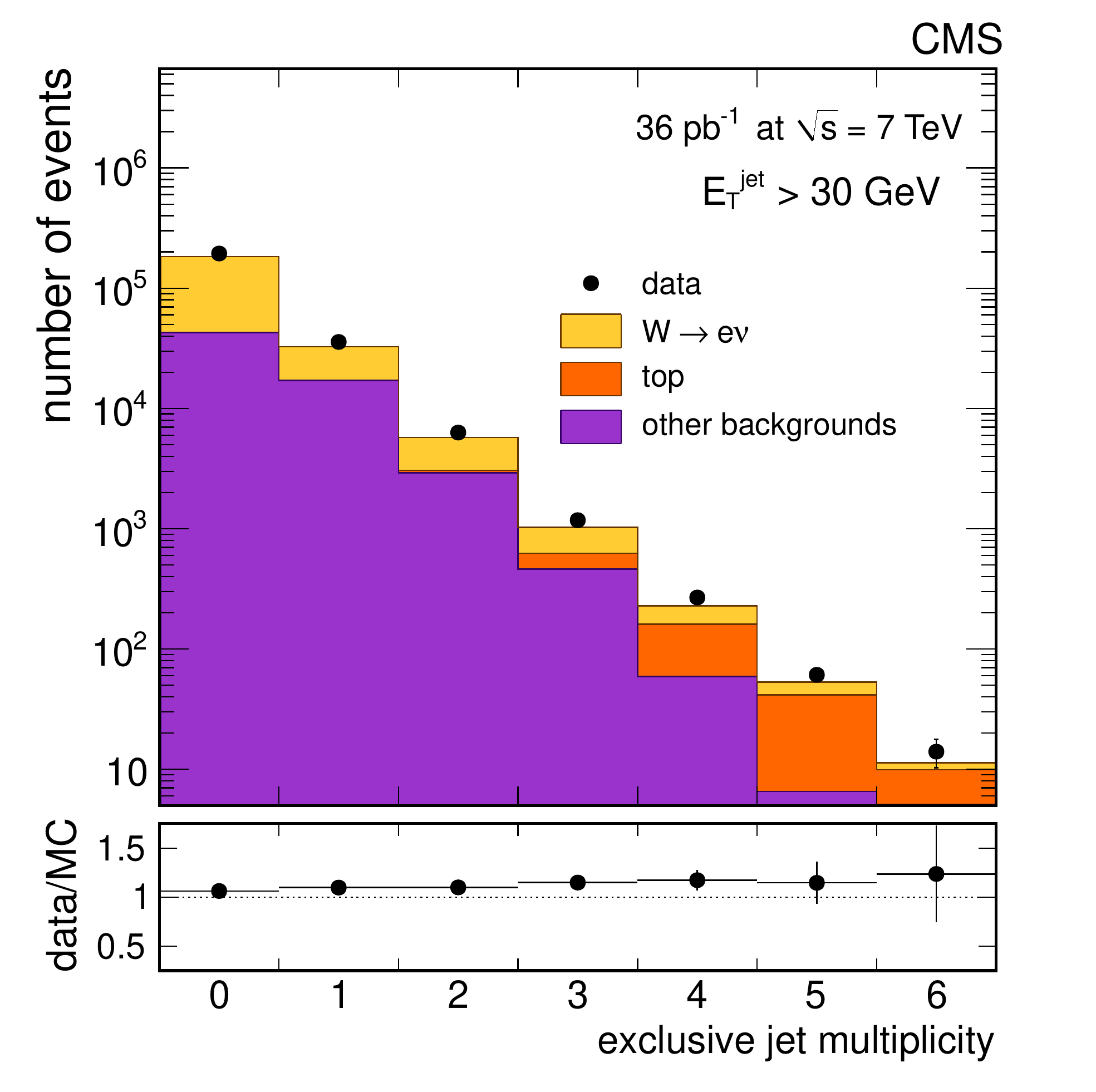}
\includegraphics[width=\sizeFigTwo]{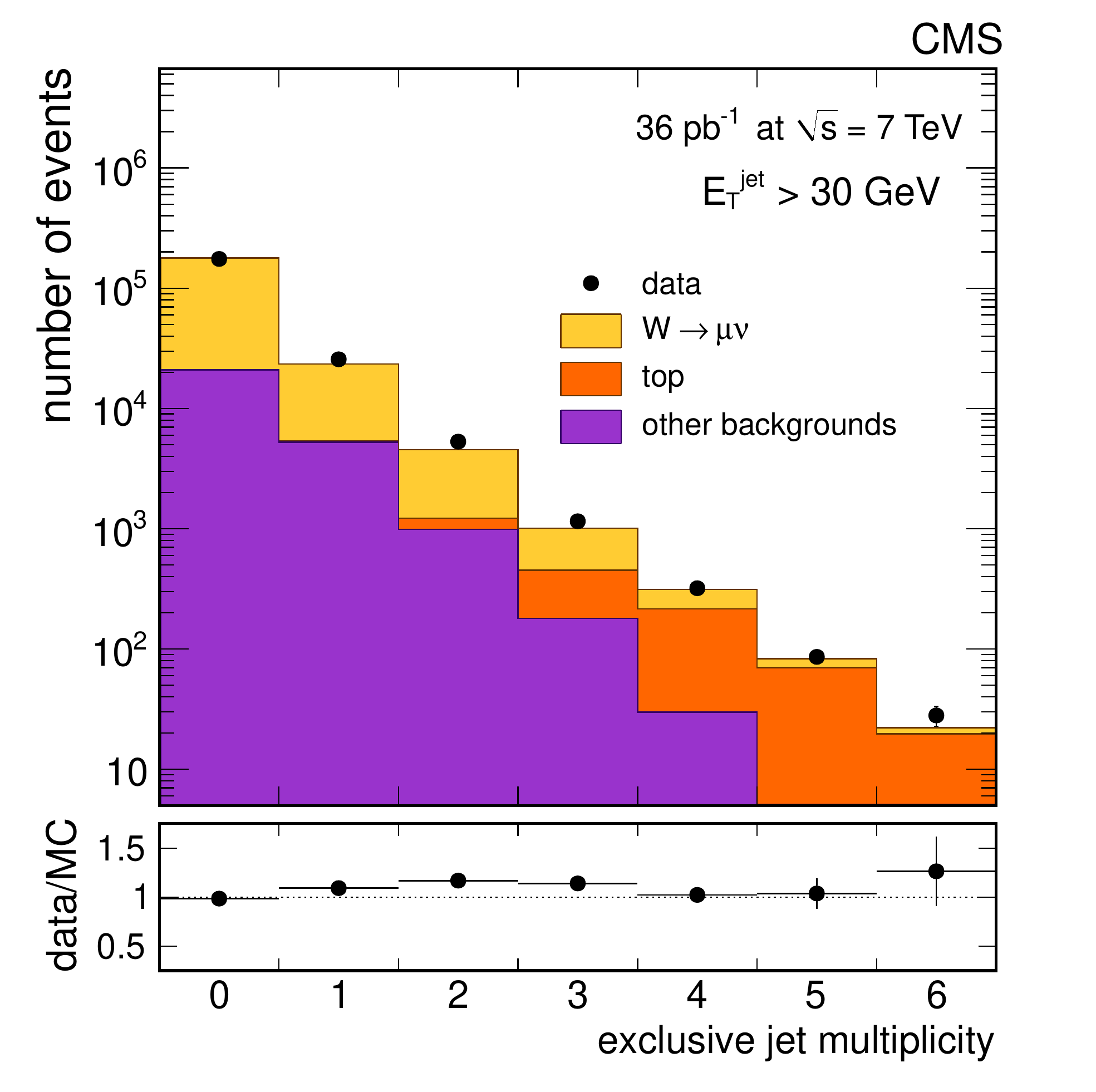}
  \caption{Exclusive number of reconstructed jets in events
    with $\Wen$ (left) and $\Wmn$ (right).  Points with error bars are
    data, histograms represent
    simulation of the signal (yellow), \ttbar and single top
    backgrounds (orange) and 
    other backgrounds (purple). Error bars represent the statistical
    uncertainty on the data. 
    These distributions have not been  corrected for detector effects
    or selection efficiency. The ratio between the data and the simulation is shown in the lower plots. 
\label{fig:Wrawrates}}
\end{figure}
\begin{figure}
  \centering
\includegraphics[width=\sizeFigTwo]{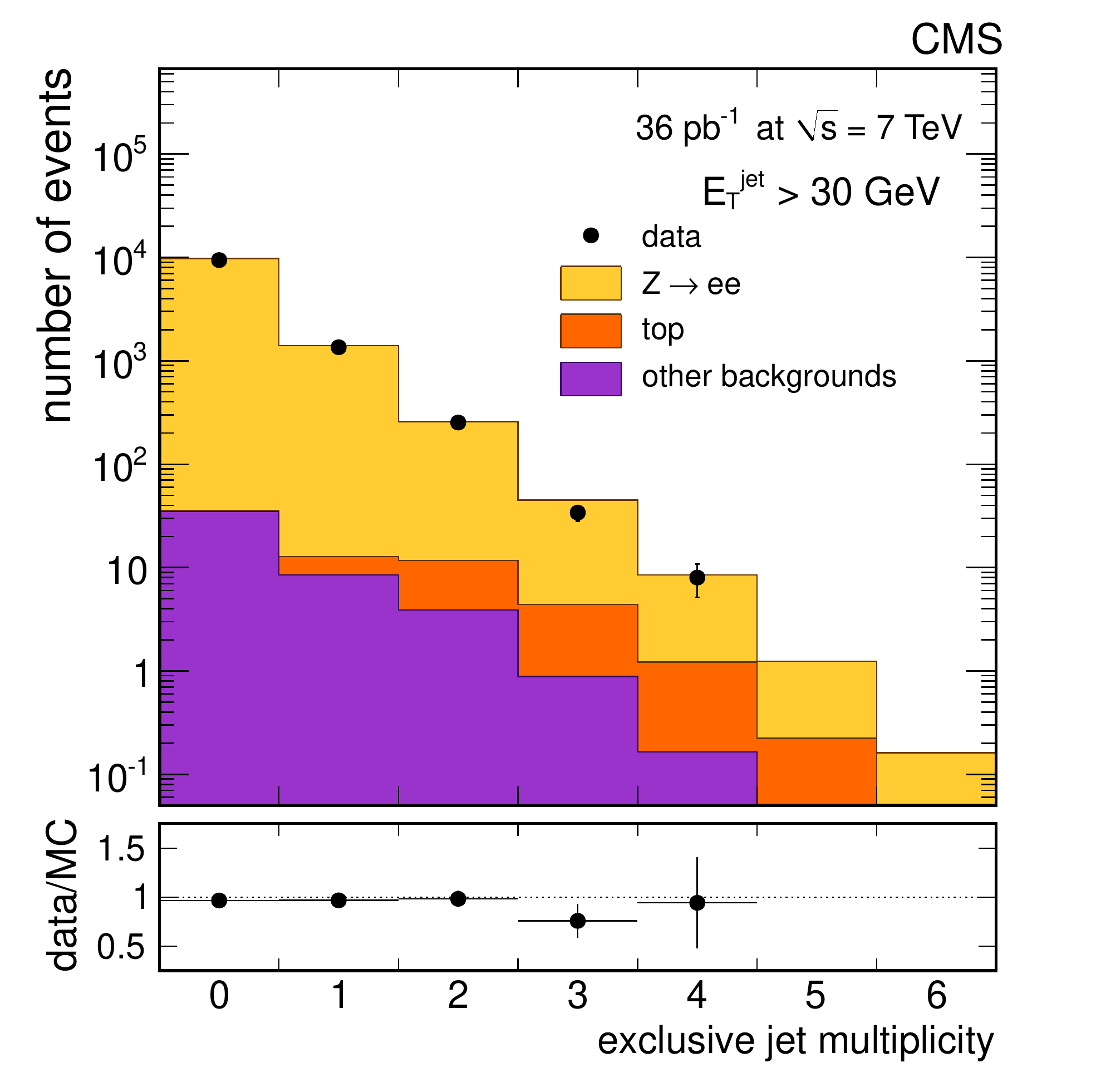}
\includegraphics[width=\sizeFigTwo]{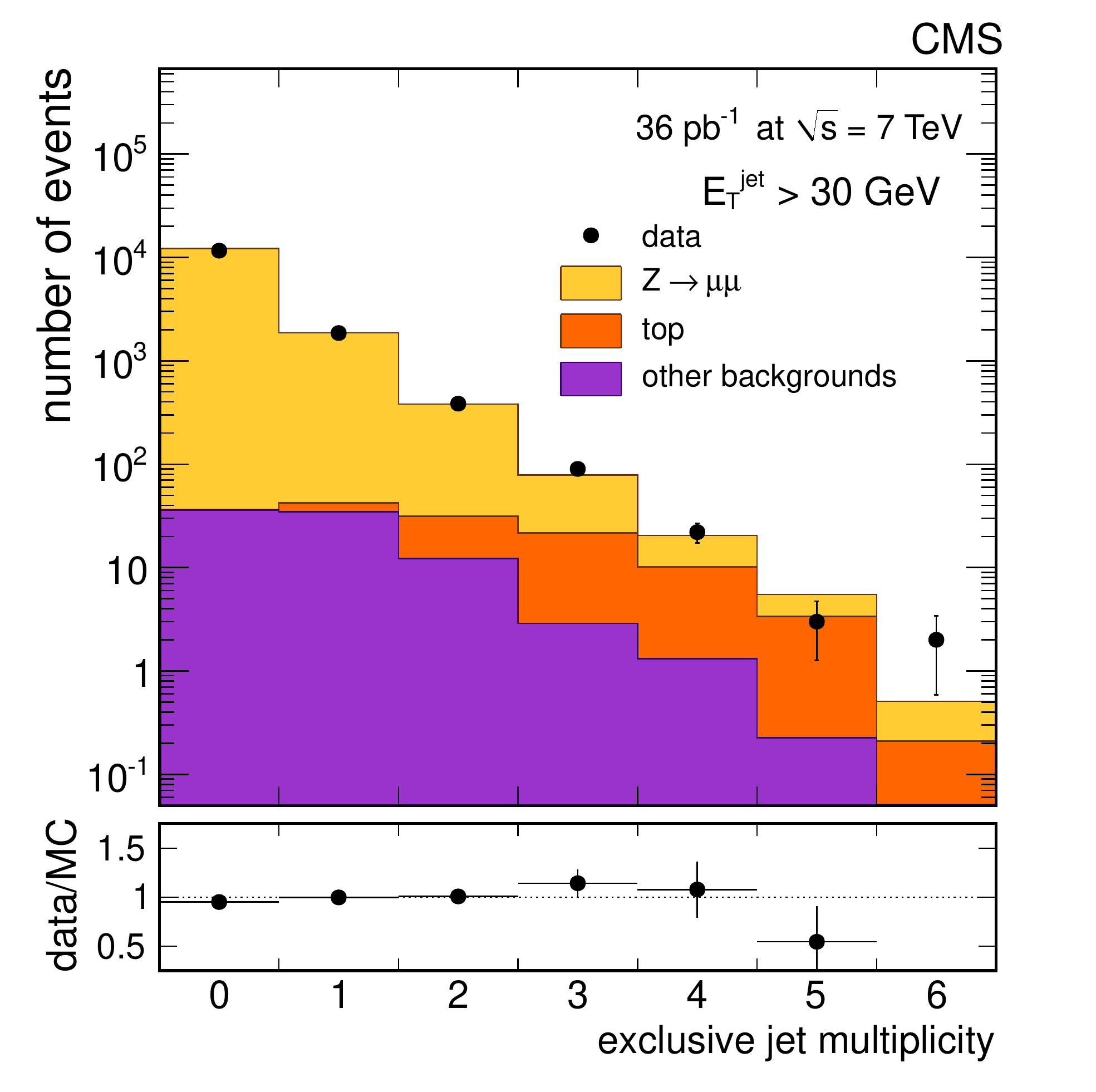}
  \caption{Exclusive number of reconstructed jets in events
    with $\Zee$ (left) and $\Zmm$ (right).  Points with error bars are data, histograms represent simulation of the signal (yellow), \ttbar and single
top backgrounds (orange) and other backgrounds (purple). Error bars represent the statistical uncertainty
on the data. These distributions have not been  corrected for detector
effects or selection efficiency. The ratio between the data and the simulation is shown in the lower plots.
\label{fig:Zrawrates}}
\end{figure}

The selected events are assigned to bins of jet
multiplicity by counting the number of jets  with \ET~$> 30$~\GeV.
The number of events with a W (Z) candidate and at least one jet are 43561
(1648) in the electron channel, and 32496 (2339) in the muon channel. 
The observed  distributions of the exclusive 
numbers of reconstructed jets in the \W\ and \Z samples are 
shown in Figs.~\ref{fig:Wrawrates} and~\ref{fig:Zrawrates}, 
respectively. The distributions from simulation are also shown. They
are normalized to the integrated luminosity using the cross-section values in
Table~\ref{tab:mcsamples}. Comparisons between 
the data and simulation distributions show that there is a very good agreement, even prior to any attempt to fit
the signal and background content of the data. 
For the \W sample, QCD multijet processes constitute the most significant source of background in the
lower jet-multiplicity bins, while top quark production dominate in the 
higher jet-multiplicity bins. For \Z events with one or more jets, the
top quark production dominates the background, which overall is very small.


\section{Acceptance and Efficiency}
\label{sec:efficiency}
\par
In order to provide model-independent results, we do not
correct for the detector acceptance, but rather quote the results 
within the acceptance, as defined by the lepton and jet fiducial
and kinematic selections given above.

\par
The efficiencies for lepton reconstruction, identification, isolation, and 
trigger are obtained from 
\zjets~data samples.
The 
sample for the measurement of a given efficiency contains events selected with two lepton 
candidates. One lepton candidate, called the  ``tag'', satisfies all  selection
requirements, including the matching to a trigger object. 
The other lepton candidate, called the ``probe'', 
is selected with criteria that depend on the efficiency being measured. The invariant mass of the tag and
probe lepton candidates must fall in the range [60--120]~\GeV.
The signal yields are obtained for two exclusive subsamples of events
in which the probe lepton passes or fails the selection criteria considered. 
Fits are performed to the invariant-mass distributions of the ``pass'' and
``fail'' subsamples, including a term that accounts for the background. 
The measured efficiency is deduced from the relative levels of signal in the ``pass'' and ``fail'' subsamples.
An estimate of the systematic uncertainty on the efficiency is obtained 
by varying the shape used to fit the Z signal among several
functional forms tested on the simulation.
The lepton selection efficiency is the product of three components: 
the reconstruction efficiency, the identification and isolation efficiency, and the trigger efficiency. 
Each of these efficiencies is calculated as a function of the jet multiplicity in the event. 
The efficiency of additional selection requirements applied to W events is computed from the simulation.
\par
For electrons we find that the efficiency is about 70\% (60\%) for the
\wjets (\zjets) signal events with variations of a few percent 
across different jet multiplicity bins, which are accounted for in the
measurement. The uncertainty ranges from about 1\% up
to about 5\%, increasing with the number of jets in the event. 
The additional requirement that  the three
methods to measure the electron charge agree, which is applied for
the charge asymmetry measurement, has an efficiency of 97\%.

\par
For muons, the efficiencies are measured as a 
function of $\PT$, $\eta$ and the jet multiplicity.
The efficiency dependence on $\PT$ and $\eta$ is measured in events
with zero and one jet. 
For events with more than one jet, where the sample size is insufficient for a
(\PT, $\eta$) dependent efficiency to be measured, an average efficiency is
computed and the (\PT, $\eta$) dependence
measured in the $n=1$ bin is assumed.  
The isolation efficiency is found to have the largest dependence on
multiplicity, while the trigger, reconstruction, and identification efficiencies
are consistent with being constant. 
The average efficiency is 85\% and 86\% for W
and Z events, respectively. They are similar because the efficiency
for selecting the second muon 
from Z candidates and for passing the \MT\ criteria with W candidates are
both near unity. While the second muon
efficiency in Z events is fairly flat across different jet
multiplicity bins, the efficiency of the \MT\ requirement for W candidates
decreases to 90\% for high jet multiplicity. The uncertainty on the
muon efficiency ranges from about 3\% up
to about 15\%, increasing with the number of jets in the event. 
\par
Electron charge misidentification is estimated with simulated samples and data.
It depends on the lepton $\eta$, with the largest
misidentification in the endcap regions.
The probability ranges from $4\times10^{-4}$ up to $3\times10^{-3}$ in the
simulation, and from $1\times10^{-3}$ up to $4\times10^{-3}$ in the data. The
difference between simulation and data is taken as uncertainty.    
The muon charge misidentification has been found to be smaller than $10^{-4}$
in cosmic-ray data~\cite{COSMICS},  therefore yielding a negligible systematic uncertainty. 
Using the tag-and-probe method, the ratio of the selection
efficiency for electrons to 
positrons is measured to equal unity within the statistical
precision of 1.4\%~\cite{Wasym}. 
The efficiency ratio between $\mu^+$
and $\mu^-$ is found equal to one within 2\%~\cite{Wasym}. 
Effects related to charge measurements
are propagated as systematic uncertainties to the final W charge asymmetry results.

\section{Signal Extraction}
\label{sec:fit}
\par
The signal yield is estimated using an extended maximum-likelihood
fit to the dilepton invariant mass ($\MLL$) distribution for the \zjetsb sample, and
to the transverse mass ($\MT$) distribution for the
\wjets sample; the number of observed events is included in the
likelihood fit as a constraint on the normalization. 
The probability distribution functions are asymmetric Gaussians with tails. Their
parameters are derived from the simulation or, for the background,  from
control data samples with inverted identification (isolation) requirement
on the electrons (muons). 
\par
For the \Z event sample, the contamination from the 
background processes, dominated by \ttbar and \wjets, is
small and does not produce a peak in the $\MLL$ distributions, 
so the $\MLL$ distribution is taken to be the sum of two components,
one for the signal and  one that accounts for all background 
processes. 
\par
For the \W sample, the background contributions can be divided 
into two components, one which exhibits a peaking structure
in $\MT$, dominated by \ttbar, and another which does not,
dominated by QCD multijet events.  We perform a two-dimensional
fit to the $\MT$ distribution and the number of b-tagged jets, $\NBJET$.
The $\MT$ distribution allows the statistical separation of
the signal from the non-peaking backgrounds, while $\NBJET$
distinguishes the signal and the background from~$\ttbar$. The
likelihood function is built under the assumption that there are no b jets in
the signal events. This implies that a fraction of \W events produced in
association with heavy-flavour jets, \ie the fraction with  
at least one heavy-flavour jet in the acceptance, is counted as
background. Considering the statistical precision of the measurement,
this assumption has a negligible effect on the \wjets cross-section
result. 
\par
\begin{figure}
  \centering
\includegraphics[width=0.38\textwidth]{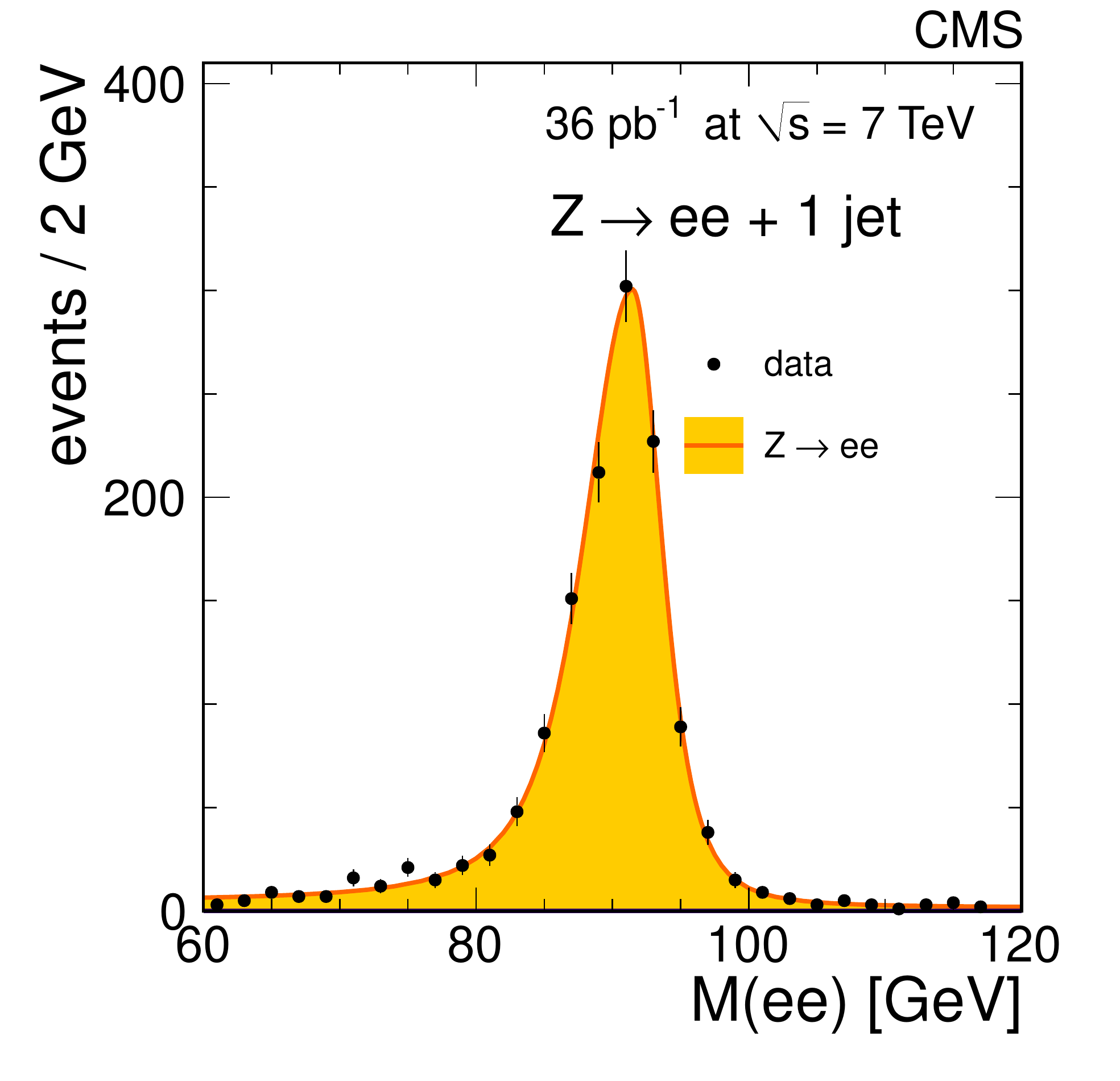}
\includegraphics[width=0.38\textwidth]{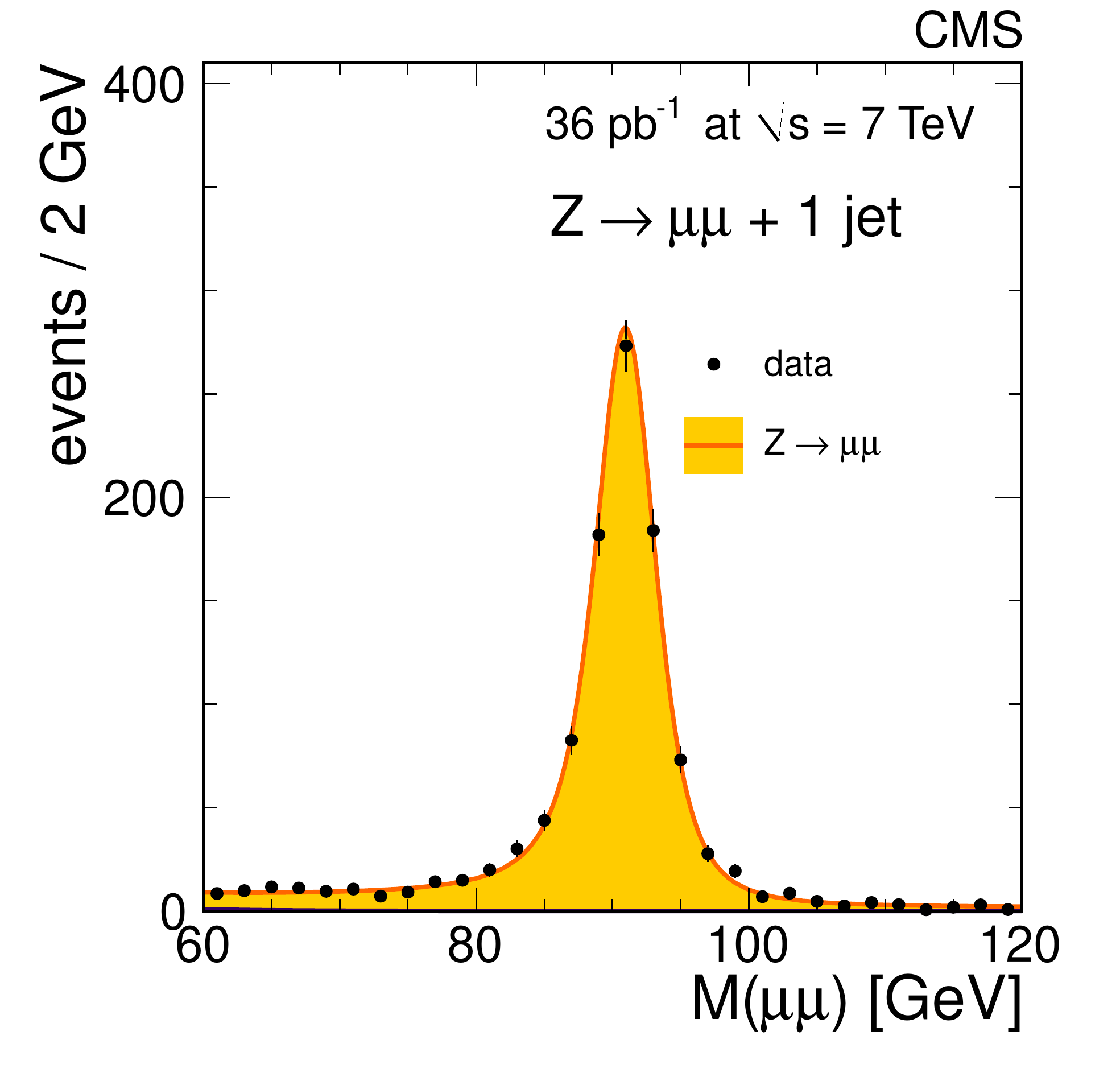}
  \caption{Dilepton mass for the $\Z+1$~jet samples, in the electron
channel (left) and the muon channel (right). Points with error bars
are data. The fit result for the signal is shown by the yellow-filled area. The background is too small to be visible in the figure. 
    \label{fig:zllfits}}
\end{figure}
\begin{figure}
  \centering
\includegraphics[width=0.45\textwidth]{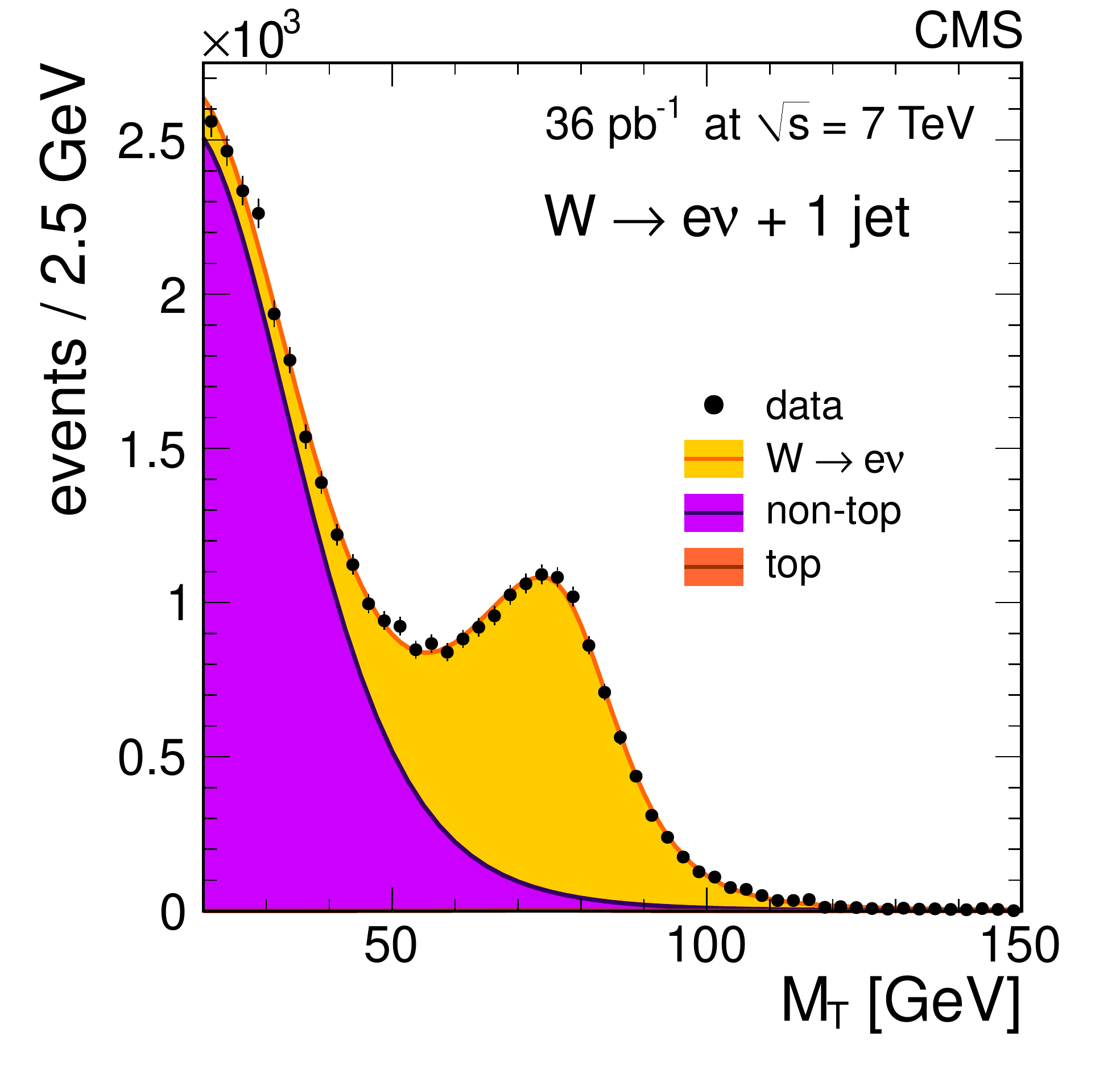}
\includegraphics[width=0.45\textwidth]{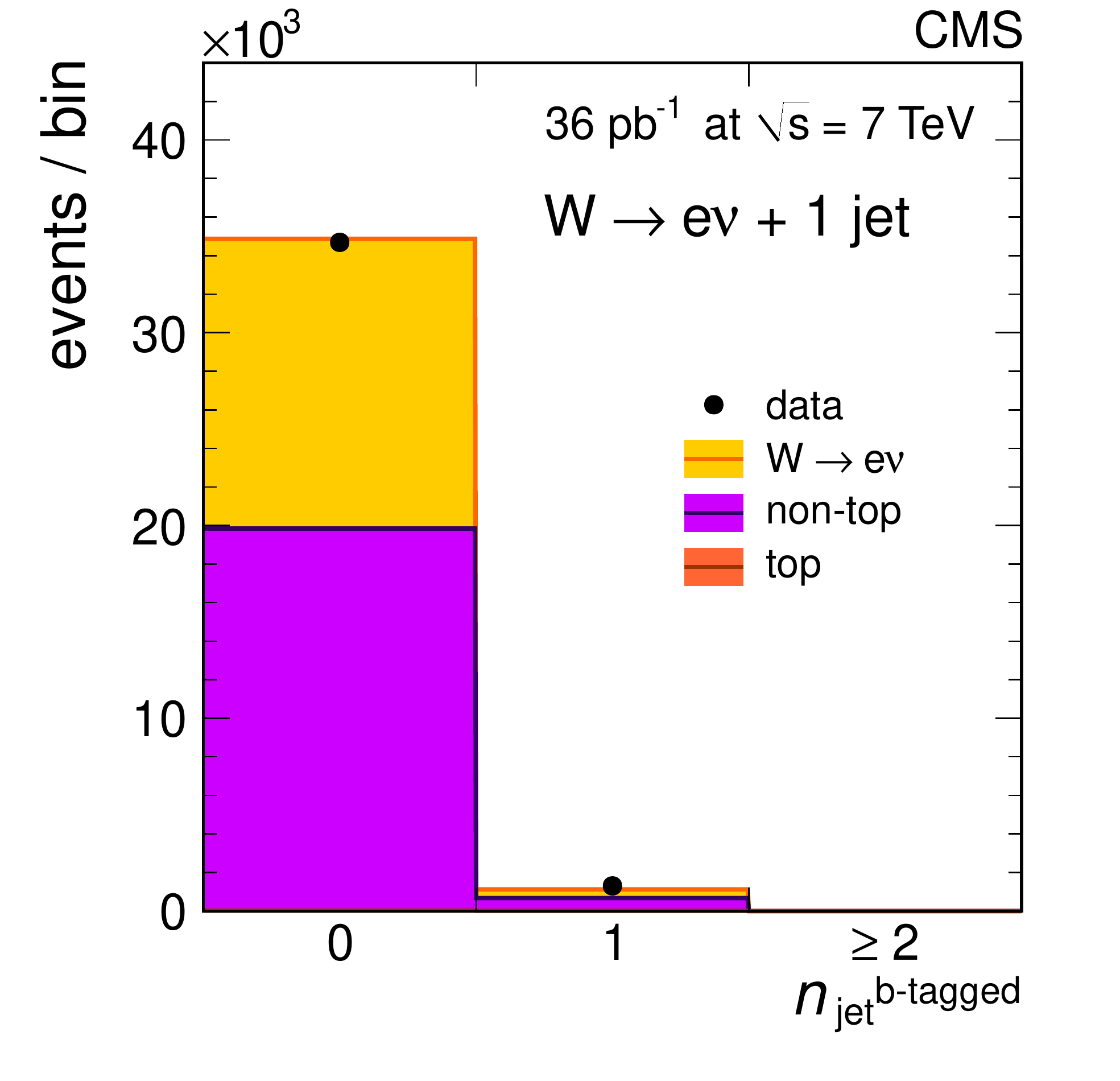}
\includegraphics[width=0.45\textwidth]{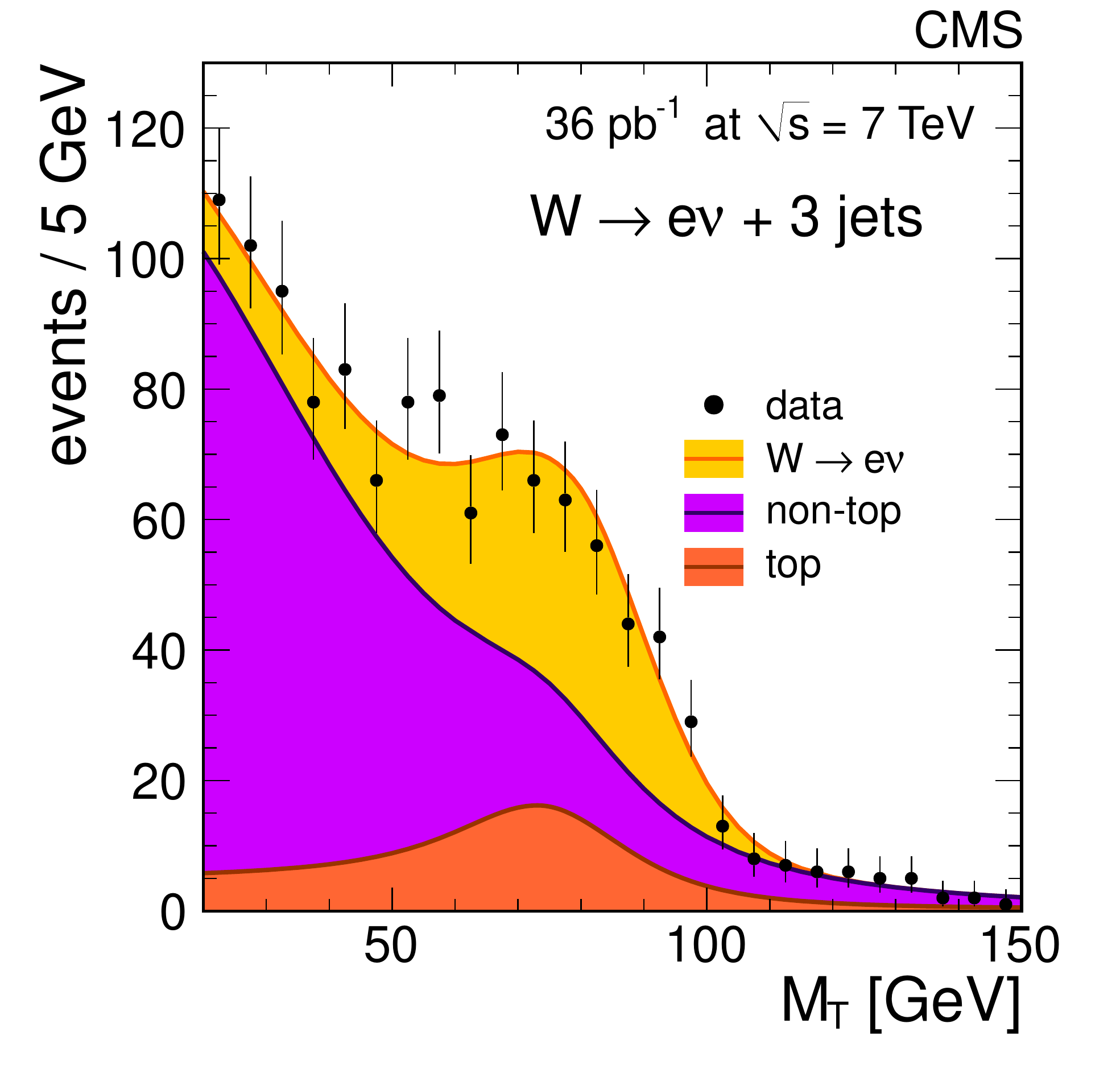}
\includegraphics[width=0.45\textwidth]{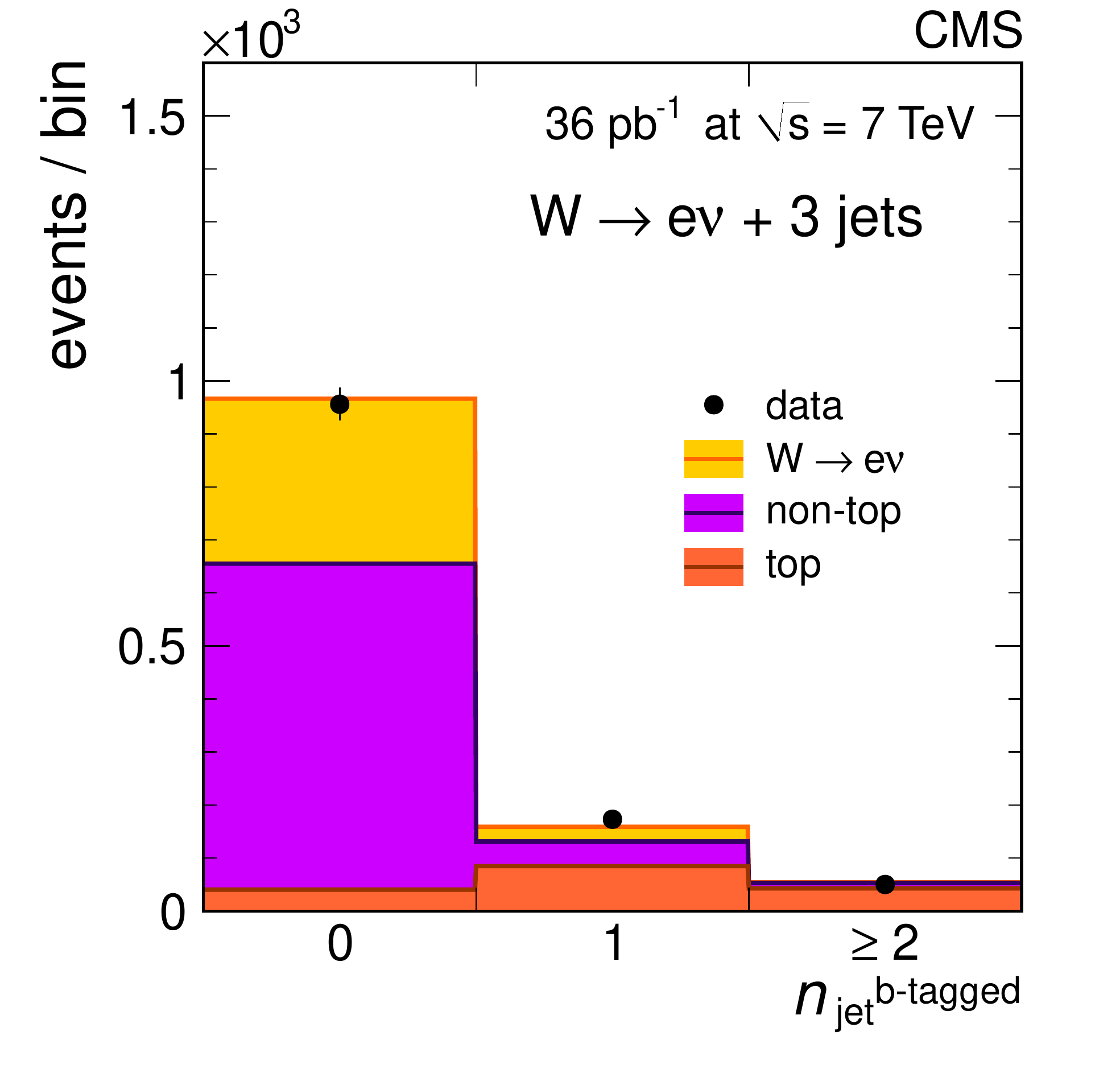}
  \caption{Fit results for the $\W(\mathrm{e}\nu)+ n$~jets sample with $n =1$ 
(upper row) and $n = 3$ (lower row). For each row, the \MT\ projection is shown on the
left, while the $\NBJET$ projection is shown on the right. Points with error bars 
are data. Fit results are shown by the colour-filled areas for the signal process (yellow),
non-top backgrounds (purple), and top backgrounds (orange).
    \label{fig:wenufits}
    }
\end{figure}
\begin{figure}
  \centering
\includegraphics[width=0.45\textwidth]{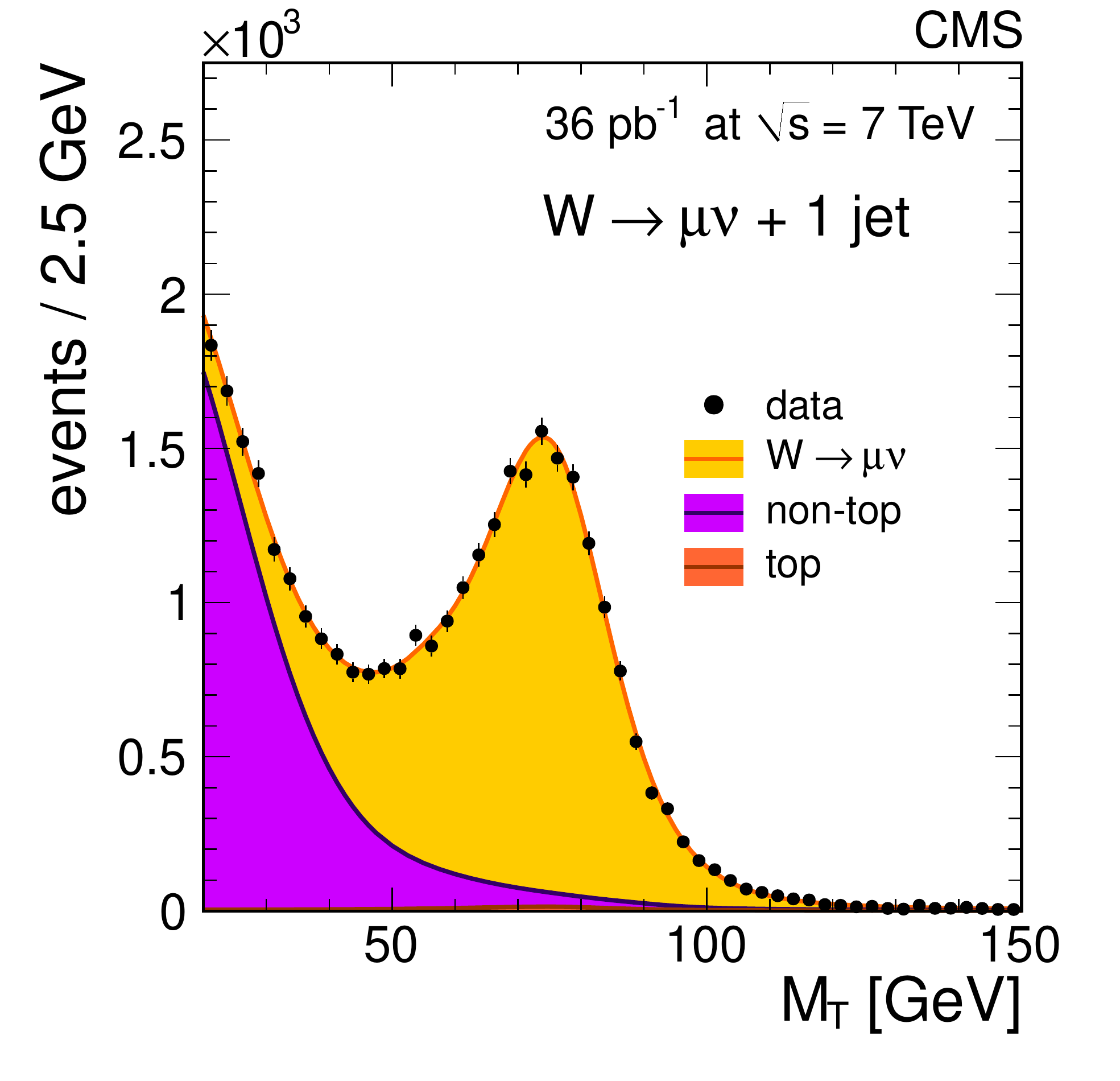}
\includegraphics[width=0.45\textwidth]{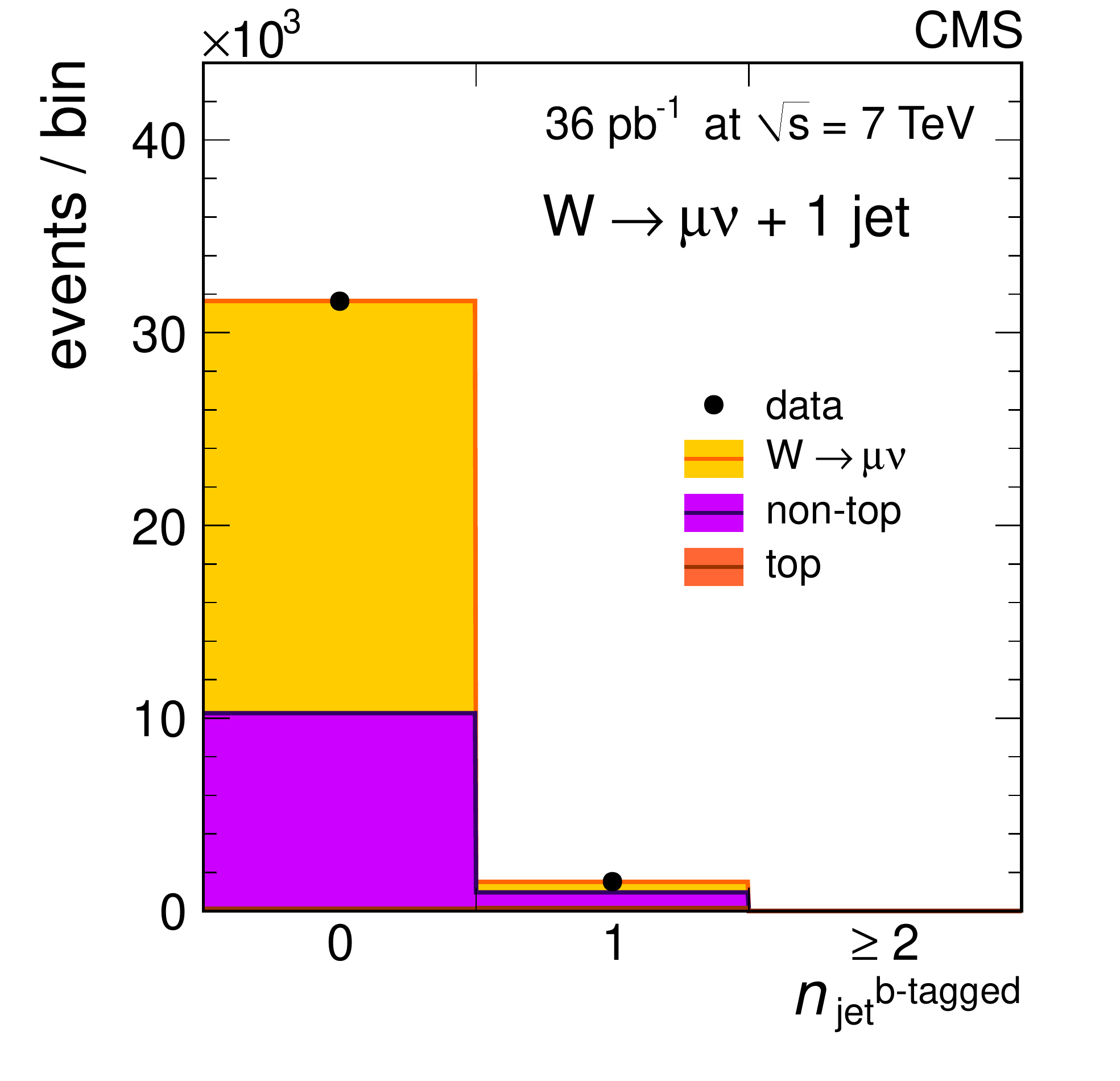}
\includegraphics[width=0.45\textwidth]{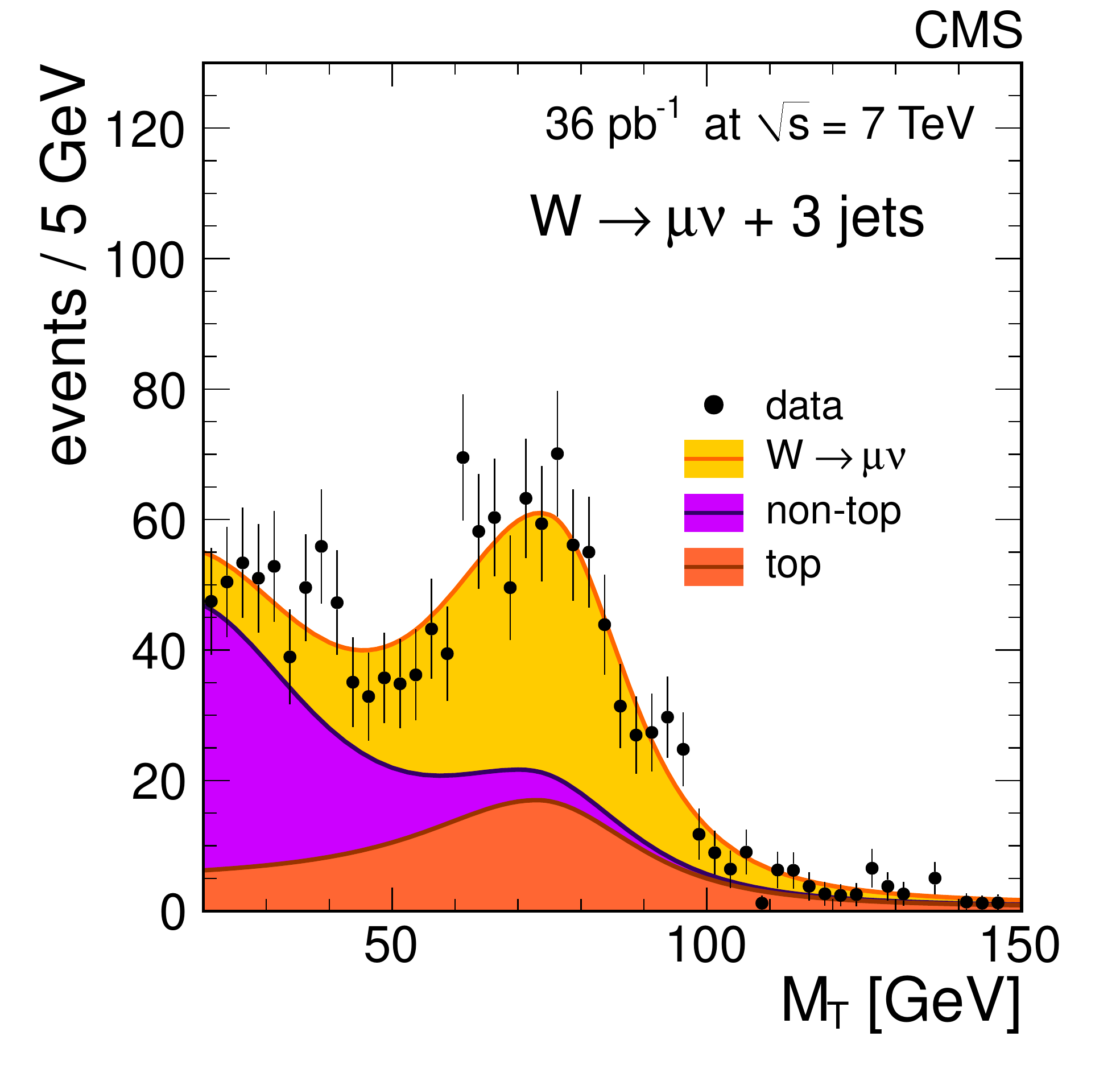}
\includegraphics[width=0.45\textwidth]{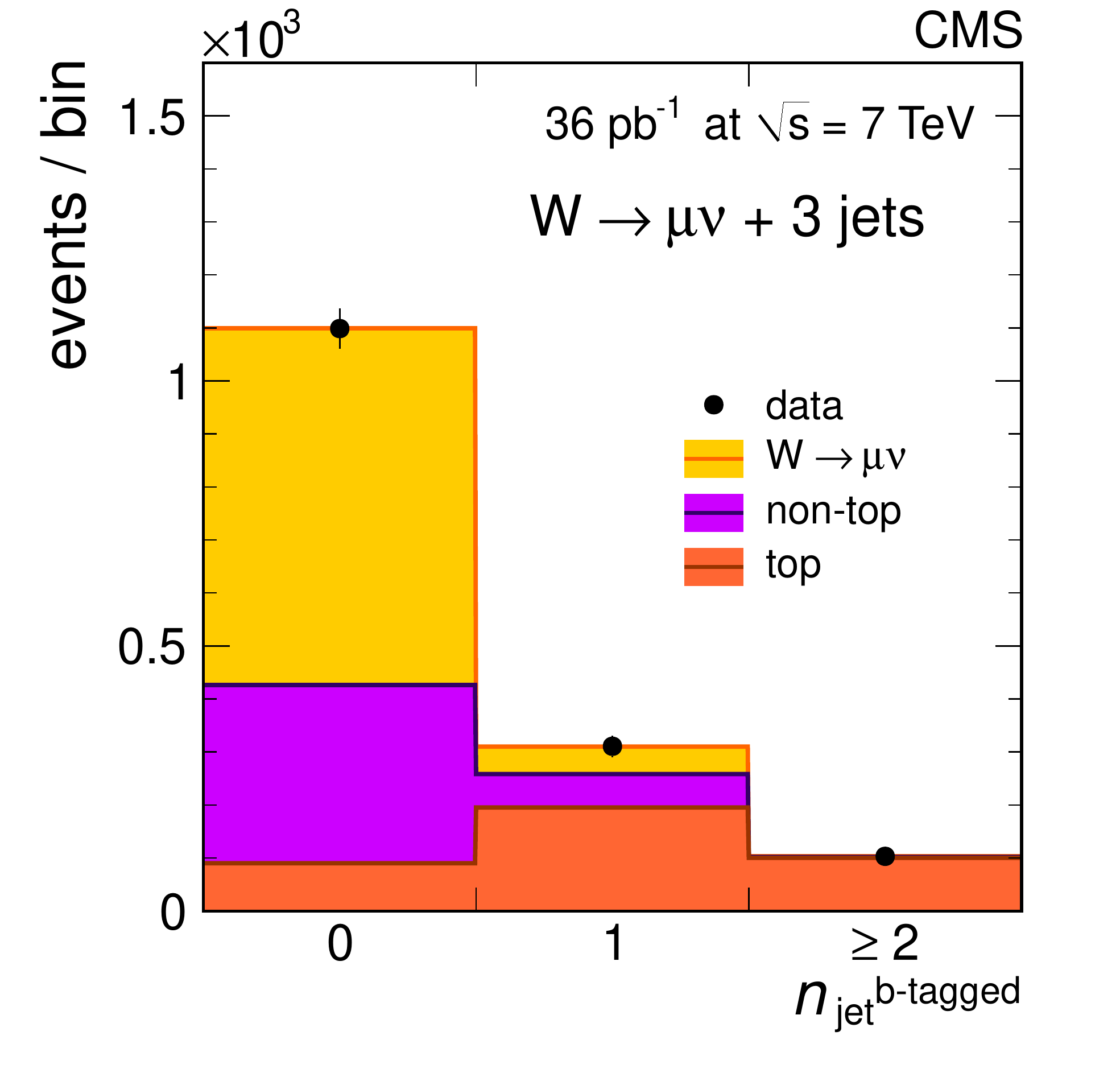}
  \caption{Fit results for the $\W(\mu\nu)+ n$~jets sample with $n =1$ 
(upper row) and $n = 3$ (lower row). For each row, the \MT\ projection is shown on the
left, while the $\NBJET$ projection is shown on the right.
Points with error bars are data. Fit results are shown by the colour-filled areas for the signal process (yellow),
non-top backgrounds (purple), and top backgrounds (orange).
    \label{fig:wmunufits}
    }
\end{figure}

The fits are done using the jet multiplicity bins for $n \le 3$; in
contrast, the jet counting is done
inclusively for the last 
bin of jet multiplicity, \ie $n \ge 4$.
Examples of fits for $\Z + 1$~jet are shown in Fig.~\ref{fig:zllfits}.
Figures~\ref{fig:wenufits} and~\ref{fig:wmunufits} show
fits in $\MT$ and $\NBJET$ projections for the $\W + n$~jets~($n$=1
and $n$=3) channel. 
The presence of the top quark background
is evident from comparing the $n = 1$ and $n = 3$
multiplicity bins.
The fit is repeated separately on the \wjetsplus and \wjetsminus samples
for the charge-asymmetry measurement.

\par
In the electron channel, the observed exclusive \vjets\ yields determined from the fit 
are corrected a posteriori for electron efficiencies, which are 
discussed in Section~\ref{sec:efficiency}.
In the muon channel, efficiencies are available in bins of the lepton 
$\PT$, $\eta$ and the jet multiplicity. This allows an
efficiency-weighted fit to be
performed, which returns efficiency-corrected yields.

\section{Unfolding}
\label{sec:unfolding}
\par
In order to estimate the scaling behaviour of the jets at the 
particle-jet level, we apply an unfolding procedure that 
removes the effects of jet energy resolution and
reconstruction efficiency.  The migration matrix, which
relates a number $n'$ of produced jets at the particle level
to an observed number $n$ of reconstructed jets, is
derived from simulated samples of \zjetsb and \wjets.
The unfolding procedure takes into account the statistical
and systematic uncertainties in the signal yields.
\par
The migration matrices are derived from simulated events with leptons and jets
within the \pt and $\eta$ acceptance of the analysis.  
The simulation is performed using 
the \MADGRAPH generator with the Z2 tune and incorporates event pileup.
\par
Two unfolding methods are employed. The baseline method is
the "singular value decomposition" (SVD) algorithm~\cite{Hocker:1995kb}.
As a crosscheck,
the iterative method~\cite{D'Agostini:1994zf} is also applied.
Both algorithms require a regularization parameter to prevent 
the statistical fluctuations in the data from appearing as
structure in the unfolded distribution. 
For the SVD method,
the regularization parameter is chosen to be $k_{\mathrm{SVD}} = 5$,
corresponding to the number of bins.  This algorithm corresponds
to an inversion of the migration matrix, and gives the most accurate uncertainty estimate. 
For the iterative algorithm, the regularization parameter $k_{\mathrm{Bayes}}=4$
is used, as suggested in Ref.~\cite{RooUnfold}.

\section{Systematic Uncertainties}
\label{sec:jetsys}
\renewcommand{\textfraction}{0}

One of the dominant sources of systematic uncertainties in the
V+jets measurements is the determination of the jet energy, which affects the jet counting.
We consider three sources of uncertainty related to the jet energy scale:

\begin{itemize}
\item Corrections applied to the measured jet energy to account for the detector response and
inhomogeneities. The corrections are derived from
measured jet \pt in dijet and photon+jet events and are available as
a function of $\eta$ and \pt~\cite{jme-10-010}.
\item The dependence of the detector response on the jet flavour. 
The difference between the flavour composition of jets in V+jets
events and the flavour composition of the  jet sample used to extract the
corrections is accounted for by an additional 2\% uncertainty on the jet energy. 
\item The average energy removed by the pileup subtraction
  method includes the activity due to the underlying event. We
  find that the jet energy is systematically decreased by
  500~\mev\ in comparison with events without pileup. This amount
  is included as a systematic uncertainty on the jet energy.
\end{itemize}

The above uncertainties on the jet energy are added in
quadrature and their effect is evaluated on the jet multiplicity distribution using
simulation. Similar results are found in all
the channels, for both W and Z events. For W events, the
effect of the mismeasured  jet energy on \MT\  
is evaluated in the fitting procedure.

Uncertainty on the jet energy resolution also affects the jet
multiplicity. The resolution is underestimated in the
simulation by $(10 \pm 10)$\% ~\cite{jme-10-014}. The
effect of this uncertainty on the jet multiplicity has been studied in simulated  \wjets 
events and  found to be below 2\%.

The pileup subtraction method was tested by comparing the jet multiplicity
in two simulated signal samples, one with pileup and one without,
with the pileup subtraction applied in the former case. The difference,
due to residual effects and jets in pileup events, is below 5\%.

All the above uncertainties are added in quadrature and the resultant
contribution to the relative systematic uncertainty on jet rates is
shown in Tables~\ref{tab:Wsyst} and \ref{tab:Zsyst}. Also shown in
these tables are the uncertainties associated with the selection
efficiency and the signal extraction procedure.
The largest source of uncertainty related to signal extraction in W events comes from the
uncertainty of the b-tagging efficiency and mistag rates. 
This uncertainty is estimated with a control sample of \ttbar\ events
decaying into e$^{\pm}\mu^{\mp}$ and a pair of b jets.  
The uncertainty due to the selection efficiency in
Tables~\ref{tab:Wsyst} and \ref{tab:Zsyst} 
includes the uncertainty on the lepton efficiency and on the selection
procedure.  
As discussed in Section~\ref{sec:efficiency}, it is evaluated with
\zjets data samples, with different strategies
for the electron and muon channels. This difference
results in a larger systematic uncertainty on the muon
channel. 
While the systematic uncertainty on the energy scale is correlated among the different jet
multiplicities, all other uncertainties are not.
The relative statistical uncertainty is also shown.

\renewcommand{\arraystretch}{1.2}

\begin{table}[b]
\caption{Relative systematic and statistical uncertainties on the measured jet multiplicity in W events, as a function of the jet multiplicity for electron and muon samples.}
\label{tab:Wsyst}
\begin{center}
\begin{tabular}{|l|r|r|r|r|r|}
\hline
\multicolumn{6}{|c|}{Uncertainties on jet rate in $\Wen$ events [\%]}\\ \hline
Jet multiplicity   & 0     &  1    & 2   & 3   & $\ge4$ \\  \hline
Energy scale and pileup & $\pm5$ & $\pm8$   &$^{+11}_{-10}$ &$^{+14}_{-12}$ &$^{+16}_{-15}$ \\
Selection efficiency  & $\pm0.5$  & $\pm0.3$    &$\pm1.0$  &$\pm1.7$ &$\pm4$ \\ 
Signal extraction  &              &  $\pm0.1$       &$\pm0.4$       &$\pm3$         &$\pm9$ \\ \hline
Total systematic uncertainty & $\pm5$  & $\pm8$    &$^{+11}_{-10}$ &$^{+14}_{-12}$ &$\pm17$ \\ \hline
Statistical uncertainty & $\pm0.3$  & $\pm1.0$    &$\pm2.4$ &$\pm10$&$\pm28$ \\
\hline
\multicolumn{6}{|c|}{Uncertainties on jet rate in $\Wmn$ events [\%]}\\ \hline 
Jet multiplicity   & 0     &  1    & 2   & 3   & $\ge4$ \\  \hline
Energy scale and pileup & $\pm5$ & $\pm8$   &$^{+11}_{-10}$ &$^{+14}_{-12}$ &$^{+16}_{-15}$ \\
Selection efficiency          & $\pm3$  & $\pm6$    &$\pm4$  &$\pm10$   &$\pm17$ \\ 
Signal extraction  &              &  $\pm0.1$       &$\pm0.4$       &$\pm3$         &$\pm9$ \\ \hline
Total systematic uncertainty & $\pm6$  & $\pm10$    &$^{+13}_{-12}$ &$^{+19}_{-17}$ &$\pm26$ \\ \hline
Statistical uncertainty & $\pm0.3$  & $\pm0.9$    &$\pm2.4$ &$\pm5.6$ &$\pm21$ \\\hline

\end{tabular}
\end{center}

\caption{Relative systematic and statistical uncertainties on the
  measured jet multiplicity rates in Z events, as a function of the
  jet multiplicity for electron and muon samples. The signal extraction
  uncertainty is not shown since it is negligible for Z events. }
\label{tab:Zsyst}
\begin{center}
\begin{tabular}{|l|r|r|r|r|r|}
\hline
\multicolumn{6}{|c|}{Uncertainties on jet rate in $\Zee$ events [\%]}\\ \hline 
Jet multiplicity   & 0     &  1    & 2   & 3   & $\ge4$ \\  \hline
Energy scale and pileup & $\pm5$ & $\pm8$   &$^{+11}_{-10}$ &$^{+14}_{-12}$ &$^{+16}_{-15}$ \\
Selection efficiency  & $\pm1.1$  & $\pm1.0$    &$\pm2.2$  &$\pm2.6$   &$\pm6$ \\ \hline
Total systematic uncertainty & $\pm5$  & $\pm8$    &$^{+11}_{-10}$ &$^{+14}_{-12}$ &$^{+17}_{-16}$ \\ \hline
Statistical uncertainty & $\pm1.1$  & $\pm3.1$    &$\pm7$ &$\pm17$ &$\pm43$ \\\hline
\multicolumn{6}{|c|}{Uncertainties on jet rate in $\Zmm$ events [\%]}\\ \hline 
Jet multiplicity   & 0     &  1    & 2   & 3   & $\ge4$ \\  \hline
Energy scale and pileup & $\pm5$ & $\pm8$   &$^{+11}_{-10}$ &$^{+14}_{-12}$ &$^{+16}_{-15}$ \\
Selection efficiency          & $\pm3$  & $^{+6}_{-5}$    &$^{+7}_{-6}$  &$\pm10$   &$^{+24}_{-12}$ \\ \hline
Total systematic uncertainty & $\pm6$  & $\pm10$    &$^{+13}_{-12}$ &$^{+18}_{-16}$ &$^{+30}_{-21}$ \\ \hline
Statistical uncertainty & $\pm0.9$  & $\pm2.6$    &$\pm5.2$ &$\pm18$ &$\pm41$ \\\hline

\end{tabular}
\end{center}
\end{table}

All statistical and systematic uncertainties are propagated in the unfolding
procedure. 
Uncertainties due to the unfolding procedure itself are calculated as
the differences between the unfolded rates using the SVD and the iterative algorithms,
and the \MADGRAPH, \PYTHIA, and Z2 or D6T tunes of \MADGRAPH for the
unfolding matrix.
The resulting uncertainties are shown with the final results in
the next section.

\section{Results}
\label{sec:results}

Results are given for the leptons within the acceptance defined by
the kinematical selection cuts in the electron (muon) channels:
leading-lepton $\PT>20$ GeV and $|\eta|<2.5$ $(2.1)$; second-lepton $\PT
> 10$ GeV and $|\eta|<2.5$ $(2.4)$. In addition, electrons in the
$1.4442 < |\eta| < 1.566$ region are excluded, and jets are not
counted if $\Delta R < 0.3$ with respect to an electron from the W or
the Z. Since the acceptance is different, we do not
combine the results for the electron and muon channels.

\begin{figure}
  \centering
  \includegraphics[width=0.55\textwidth]{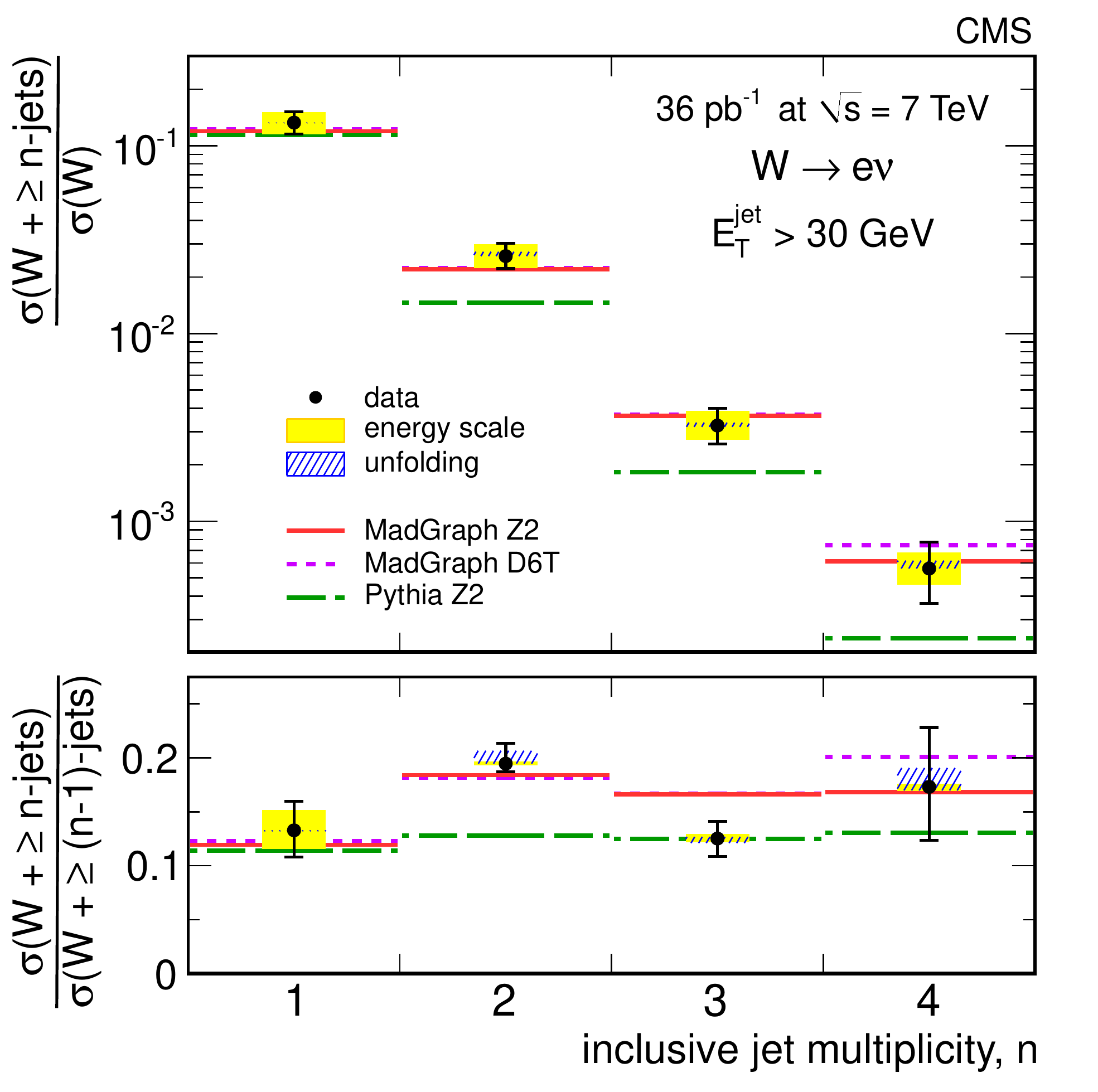}
  \caption{
The ratios $\sigma(\W+\ge n~{\mathrm{jets}})/\sigma(\W)$ (top) and
$\sigma(\W+\ge n~{\mathrm{jets}}) / \sigma(\W+\ge
(n-1)~{\mathrm{jets}})$ (bottom) in the electron channel
compared with the expectations from two \MADGRAPH tunes and
\PYTHIA. Points with error bars correspond to the data. The
uncertainties due to the energy scale and unfolding procedure are
shown as yellow and hatched bands, respectively. The error bars
represent the total uncertainty.
\label{fig:Wenurates}}
\end{figure}

\begin{figure}
  \centering
  \includegraphics[width=0.55\textwidth]{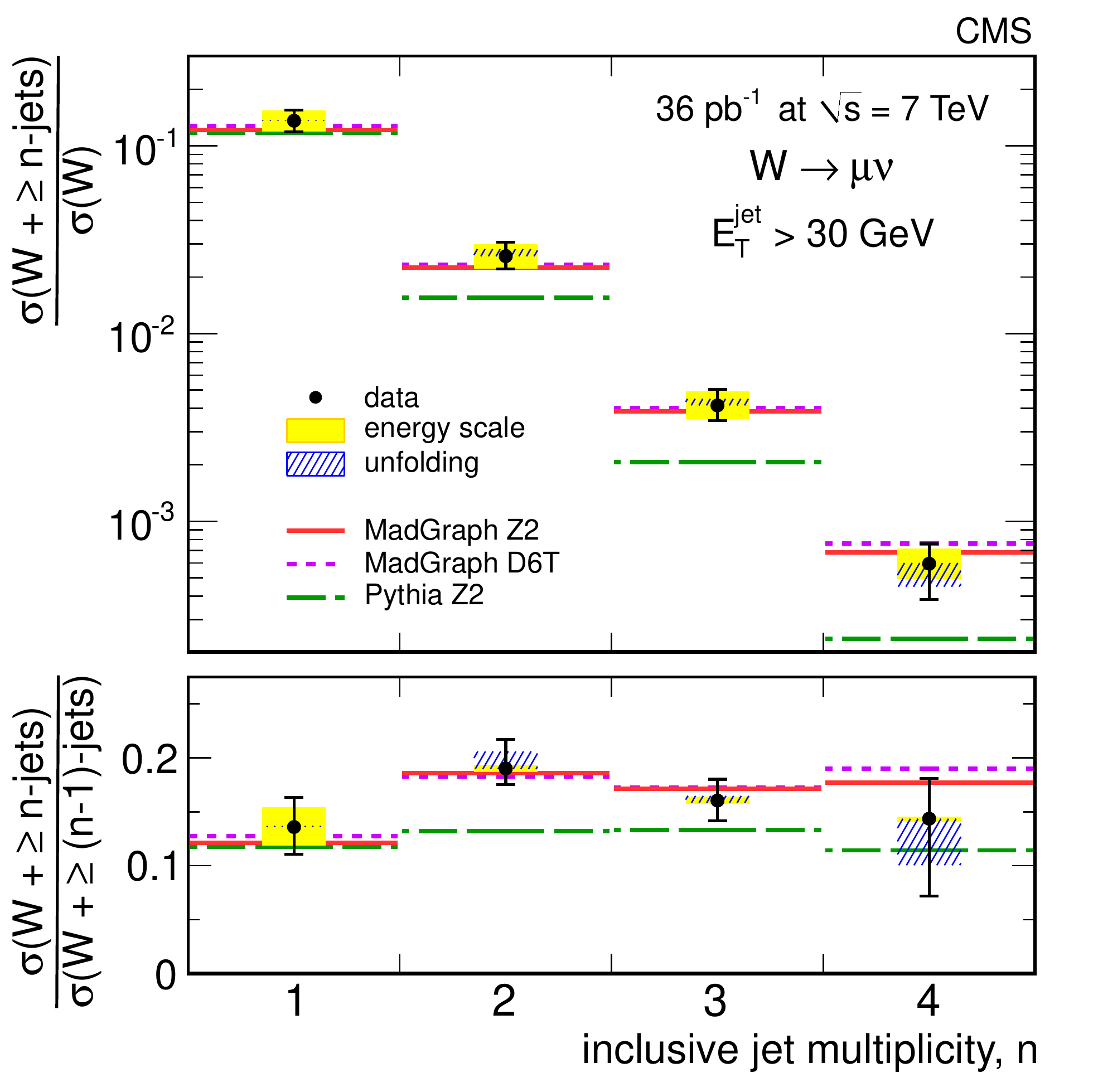}
  \caption{
The ratio  $\sigma(\W+\ge n~{\mathrm{jets}})/\sigma(\W)$ (top) and
$\sigma(\W+\ge n~{\mathrm{jets}}) / \sigma(\W+\ge
(n-1)~{\mathrm{jets}})$ (bottom)
in the muon channel
compared with the expectations from two \MADGRAPH tunes and
\PYTHIA. Points with error bars correspond to the data. The
uncertainties due to the energy scale and unfolding procedure are
shown as yellow and hatched bands, respectively. The error bars
represent the total uncertainty.
\label{fig:Wmunurates}}
\end{figure}

\begin{figure}
  \centering
  \includegraphics[width=0.55\textwidth]{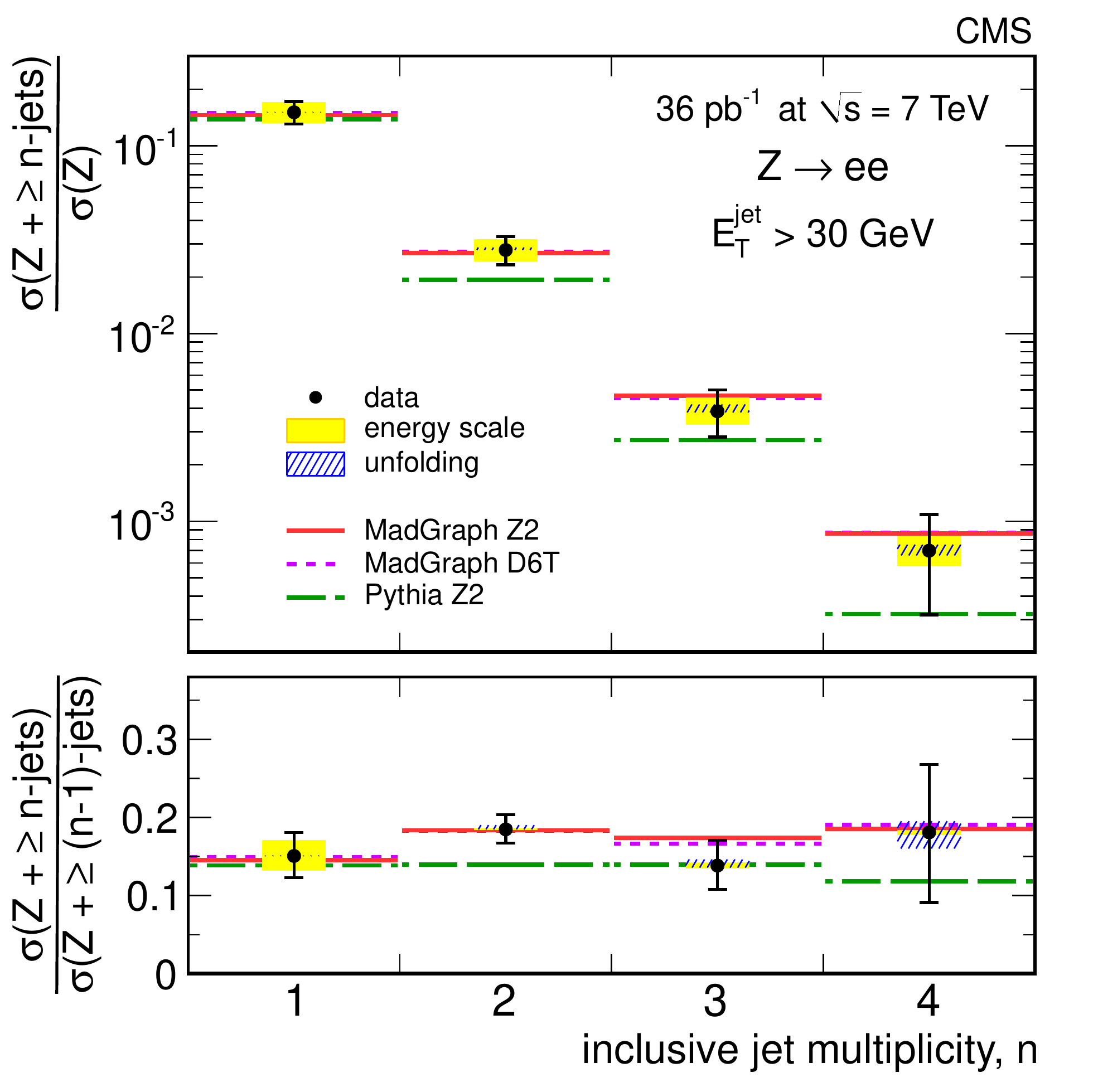}
  \caption{
The ratio  $\sigma(\Z+\ge n~{\mathrm{jets}})/\sigma(\Z)$ (top) and
$\sigma(\Z+\ge n~{\mathrm{jets}}) / \sigma(\Z+\ge
(n-1)~{\mathrm{jets}})$ (bottom) in the electron channel
compared with the expectations from two \MADGRAPH tunes and
\PYTHIA. Points with error bars correspond to the data. The
uncertainties due to the energy scale and unfolding procedure are
shown as yellow and hatched bands, respectively. The error bars
represent the total uncertainty.
\label{fig:Zeerates}}
\end{figure}

\begin{figure}
  \centering
  \includegraphics[width=0.55\textwidth]{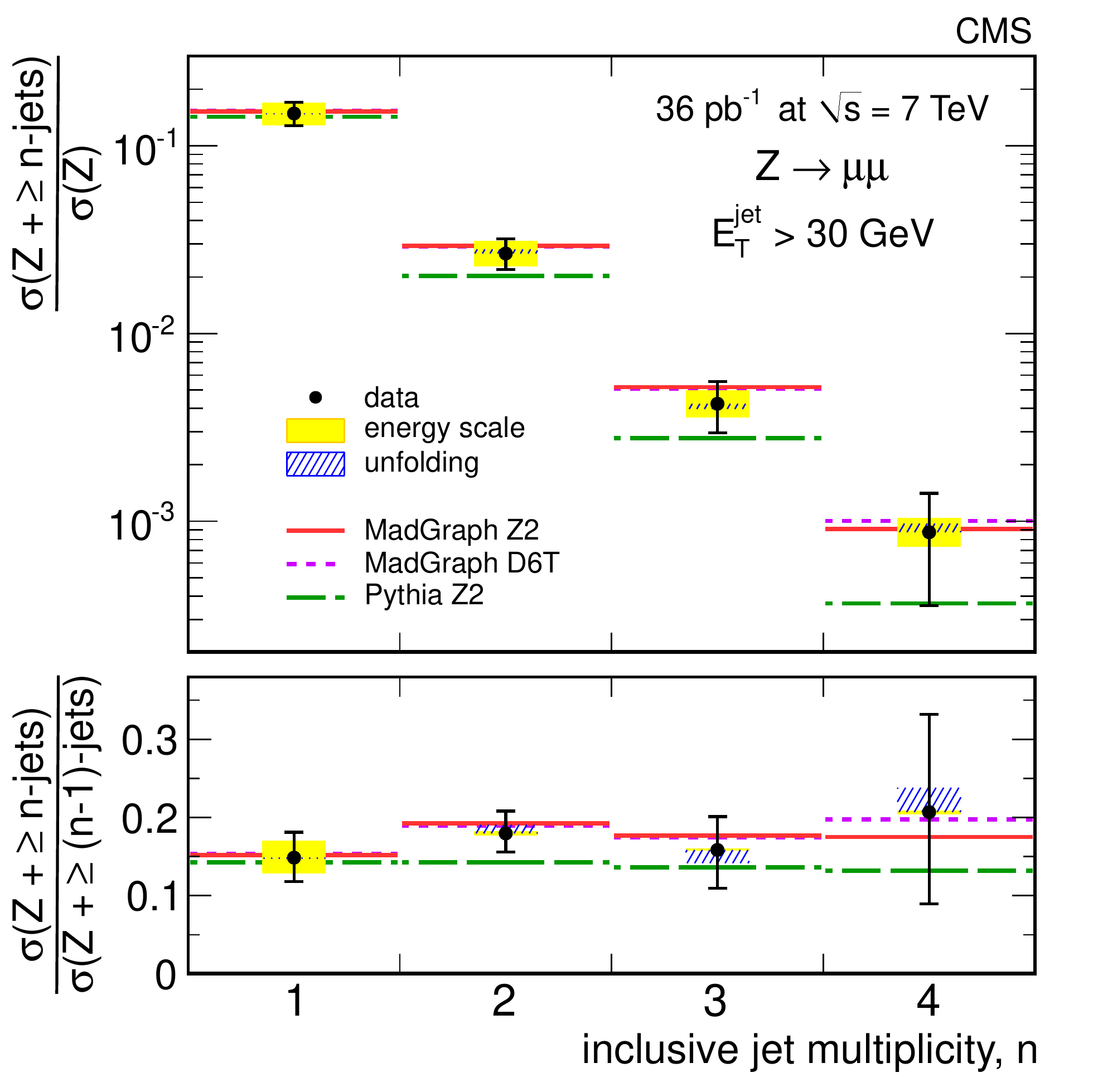}
  \caption{
The ratio  $\sigma(\Z+\ge n~{\mathrm{jets}})/\sigma(\Z)$ (top) and
$\sigma(\Z+\ge n~{\mathrm{jets}}) / \sigma(\Z+\ge
(n-1)~{\mathrm{jets}})$ (bottom) in the muon channel
compared with the expectations from two \MADGRAPH tunes and
\PYTHIA. Points with error bars correspond to the data. The
uncertainties due to the energy scale and unfolding procedure are
shown as yellow and hatched bands, respectively. The error bars
represent the total uncertainty.
\label{fig:Zmumurates}}
\end{figure}

Two sets of ratios from the unfolded jet multiplicity distributions are calculated
The first set of ratios is $\sigma(\PV+\ge n~{\mathrm{jets}})/\sigma(\PV)$,
where $\sigma(\PV)$ is the inclusive cross section, and is presented
in Tables~\ref{tab:inclusiveRatiosNNtotW} and~\ref{tab:inclusiveRatiosNNtotZ}
and in the upper frames of Figs.~\ref{fig:Wenurates}--\ref{fig:Zmumurates}.
The second set of ratios is
$\sigma(\PV+\ge n~{\mathrm{jets}})/\sigma(\PV+\ge (n-1)~{\mathrm{jets}})$,
reported in Tables~\ref{tab:inclusiveRatiosN1NW}
and~\ref{tab:inclusiveRatiosN1NZ} and in the lower frames of
Figs.~\ref{fig:Wenurates}--\ref{fig:Zmumurates}.
The contributions of the systematic uncertainties associated with the
jet energy scale (including pileup effects) and the unfolding are
given in the tables and shown as error bands in the
figures.
For $n\ge2$ jets, the \PYTHIA pure parton shower
simulation fails to describe the data, while the
\MADGRAPH simulation agrees well with the experimental spectrum.
Because of the jet threshold $\ET > 30$~GeV, the sensitivity
to the tuning of the underlying event is negligible. The expectations from
the simulation with the Z2 and the D6T tunes are identical. Thus, in the
rest of the paper we will only consider the Z2 tune for comparison
with the data.

The statistical uncertainty quoted
in the third column of Tables~\ref{tab:inclusiveRatiosNNtotW}--\ref{tab:inclusiveRatiosN1NZ}
includes only the statistical contribution from the fit results. It is combined (fourth
column) with the systematic uncertainties from the fit and the
statistical and systematic uncertainty on the efficiency, which are
uncorrelated between the samples with different numbers of jets.
To estimate the jet energy scale uncertainty,
jet rates from the fits were scaled higher and lower according
to this uncertainty. Those numbers were then unfolded
and the difference in the output from the actual fit value is quoted
in the fifth column. Finally, the uncertainty due to the unfolding
algorithm is evaluated as explained above and is shown in
the last column of the tables.

\begin{table}[b]
\caption{ Results for
$\sigma(W+\ge n~{\mathrm{jets}}) / \sigma(W)$ in the electron and muon
channels. A full description of the uncertainties is given in the text.
\label{tab:inclusiveRatiosNNtotW}
}
\begin{center}
{  
\begin{tabular}{ | c | c | c | c | c | c | }
\hline
 $n$ jets & $\frac{\sigma(W+\ge n~{\mathrm{jets}})}{\sigma(W)}$ & stat. &  stat. + efficiency      & energy      & unfolding\\
          &                &       &  and signal extraction   & scale &          \\
\hline\hline
\multicolumn{6}{|c|}{electron channel} \\
\hline
$\geq$ 1  jets        &  0.133        &  0.002        &  0.002        &  $^{+ 0.019 }_{-0.017}$       &  $\pm 0.001 $       \\
$\geq$ 2  jets        &  0.026        &  0.001        &  0.001        &  $\pm 0.004$               &  $\pm 0.001 $       \\
$\geq$ 3  jets        &  0.0032       &  0.0004       &  0.0004       &  $^{+ 0.0006 }_{-0.0005}$     &  $\pm 0.0001 $     \\
$\geq$ 4  jets        &  0.00056      &  0.00017      &  0.00018      &  $^{+ 0.00012 }_{-0.00010}$   &  $^{+ 0.00006 }_{-0.00001}$   \\
\hline
\multicolumn{6}{|c|}{muon channel} \\
\hline
$\geq$ 1  jets        &  0.136        &  0.002        &  0.007        &  $^{+ 0.019 }_{-0.017}$       &  $\pm 0.001$       \\
$\geq$ 2  jets        &  0.026        &  0.001        &  0.002        &  $\pm 0.004$               &  $^{+ 0.002 }_{-0.001}$       \\
$\geq$ 3  jets        &  0.0041       &  0.0003       &  0.0005       &  $^{+ 0.0008 }_{-0.0006}$     &  $^{+ 0.0003 }_{-0.0001}$     \\
$\geq$ 4  jets        &  0.00059      &  0.00011      &  0.00017      &  $^{+ 0.00012 }_{-0.00010}$   &  $^{+ 0.00001 }_{-0.00015}$   \\
\hline
\end{tabular}
}
\end{center}

\caption{ Results for
$\sigma(Z+\ge n~{\mathrm{jets}}) / \sigma(Z)$ in the electron and muon
channels.  A full description of the uncertainties is given in the text.
\label{tab:inclusiveRatiosNNtotZ}
}
\begin{center}
{  
\begin{tabular}{ | c | c | c | c | c | c | }
\hline
 $n$ jets &  $\frac{\sigma(Z+\ge n~{\mathrm{jets}})}{\sigma(Z)}$  & stat. &  stat. + efficiency    & energy      & unfolding\\
          &                &       & and signal extraction  & scale &          \\
\hline\hline
\multicolumn{6}{|c|}{electron channel} \\
\hline
$\geq$ 1  jets        &  0.151        &  0.006        &  0.006        &  $^{+ 0.021 }_{-0.019}$       &  $\pm 0.001$       \\
$\geq$ 2  jets        &  0.028        &  0.003        &  0.003        &  $\pm 0.004$                &  $\pm 0.001$       \\
$\geq$ 3  jets        &  0.0039       &  0.0009       &  0.0009       &  $^{+ 0.0007 }_{-0.0006}$     &  $^{+ 0.0003 }_{-0.0001}$     \\
$\geq$ 4  jets        &  0.00070      &  0.00036      &  0.00036      &  $^{+ 0.00014 }_{-0.00012}$   &  $^{+ 0.00005 }_{-0.00004}$   \\
\hline\hline
\multicolumn{6}{|c|}{muon channel} \\
\hline
$\geq$ 1  jets        &  0.149        &  0.005        &  0.011        &  $^{+ 0.022 }_{-0.020}$       &  $\pm 0.001$       \\
$\geq$ 2  jets        &  0.027        &  0.003        &  0.004        &  $\pm 0.004$               & $\pm 0.001$       \\
$\geq$ 3  jets        &  0.0042       &  0.0011       &  0.0012       &  $^{+ 0.0008 }_{-0.0006}$     &  $^{+ 0.0001 }_{-0.0003}$     \\
$\geq$ 4  jets        &  0.00087      &  0.00050      &  0.00056      &  $^{+ 0.00017 }_{-0.00015}$   &  $^{+ 0.00010 }_{-0.00001}$   \\
\hline
\end{tabular}
}
\end{center}
\end{table}

\begin{table}[b]
\caption{Results for
$\sigma(W+\ge n~{\mathrm{jets}}) / \sigma(W+\ge (n-1)~{\mathrm{jets}})$ in the electron and muon
channels.  A full description of the uncertainties is given in the text.
\label{tab:inclusiveRatiosN1NW}
}
\begin{center}
{  
\begin{tabular}{ | c | c | c | c | c | c | }
\hline
 $n$ jets &  $\frac{\sigma(W+\ge n~{\mathrm{jets}})}{\sigma(W+\ge (n-1)~{\mathrm{jets}})}$  & stat. &  stat. + efficiency   & energy      & unfolding\\
          &                &       & and signal extraction & scale &          \\
\hline\hline
\multicolumn{6}{|c|}{electron channel} \\
\hline
$\geq$ 1 / $\geq$ 0 jets 	&  0.133  	&  0.002  	&  0.002  	&  $^{+ 0.019 }_{-0.017}$  	&  $\pm 0.001$  	        \\
$\geq$ 2 / $\geq$ 1 jets 	&  0.195  	&  0.007  	&  0.007  	&  $^{+ 0.002 }_{-0.001}$  	&  $^{+ 0.012 }_{-0.001}$  	\\
$\geq$ 3 / $\geq$ 2 jets 	&  0.125  	&  0.014  	&  0.015  	&  $\pm 0.004$          	&  $^{+ 0.002 }_{-0.004}$  	\\
$\geq$ 4 / $\geq$ 3 jets 	&  0.173  	&  0.046  	&  0.049  	&  $^{+ 0.003 }_{-0.004}$  	&  $^{+ 0.017 }_{-0.003}$  	\\
\hline\hline
\multicolumn{6}{|c|}{muon channel} \\
\hline
$\geq$ 1 / $\geq$ 0 jets 	&  0.136  	&  0.002  	&  0.007  	&  $^{+ 0.019 }_{-0.017}$  	&  $\pm 0.001$         	 \\
$\geq$ 2 / $\geq$ 1 jets 	&  0.190  	&  0.005  	&  0.014  	&  $^{+ 0.004 }_{-0.003}$  	&  $^{+ 0.016 }_{-0.001}$  	\\
$\geq$ 3 / $\geq$ 2 jets 	&  0.160  	&  0.011  	&  0.018  	&  $^{+ 0.004 }_{-0.003}$  	&  $^{+ 0.004 }_{-0.002}$  	\\
$\geq$ 4 / $\geq$ 3 jets 	&  0.144  	&  0.025  	&  0.037  	&  $^{+ 0.002 }_{-0.003}$  	&  $^{+ 0.001 }_{-0.043}$  	\\
\hline
\end{tabular}
}
\end{center}

\caption{ Results for
$\sigma(Z+\ge n~{\mathrm{jets}}) / \sigma(Z+\ge (n-1)~{\mathrm{jets}})$ in the electron and muon
channels.  A full description of the uncertainties is given in the text.
\label{tab:inclusiveRatiosN1NZ}
}
\begin{center}
{
\begin{tabular}{ | c | c | c | c | c | c | }
\hline
 $n$ jets & $\frac{\sigma(Z+\ge n~{\mathrm{jets}})}{\sigma(Z+\ge (n-1)~{\mathrm{jets}})}$ & stat. &  stat. + efficiency   & energy     & unfolding\\
          &                &       & and signal extraction & scale &          \\
\hline\hline
\multicolumn{6}{|c|}{electron channel} \\
\hline
$\geq$ 1 / $\geq$ 0 jets 	&  0.151  	&  0.006  	&  0.006  	&  $^{+ 0.021 }_{-0.019}$  	&  $\pm 0.001$  	        \\
$\geq$ 2 / $\geq$ 1 jets 	&  0.185  	&  0.017  	&  0.017  	&  $^{+ 0.002 }_{-0.001}$  	&  $^{+ 0.006 }_{-0.001}$  	\\
$\geq$ 3 / $\geq$ 2 jets 	&  0.138  	&  0.030  	&  0.030  	&  $\pm 0.004$         	&  $^{+ 0.008 }_{-0.003}$  	\\
$\geq$ 4 / $\geq$ 3 jets 	&  0.181  	&  0.085  	&  0.085  	&  $^{+ 0.003 }_{-0.004}$  	&  $^{+ 0.014 }_{-0.021}$  	\\

\hline\hline
\multicolumn{6}{|c|}{muon channel} \\
\hline
$\geq$ 1 / $\geq$ 0 jets 	&  0.149  	&  0.005  	&  0.011  	&  $^{+ 0.022 }_{-0.020}$  	&  $\pm 0.001$         	 \\
$\geq$ 2 / $\geq$ 1 jets 	&  0.180  	&  0.016  	&  0.023  	&  $\pm 0.003$  	        &  $^{+ 0.011 }_{-0.001}$  	\\
$\geq$ 3 / $\geq$ 2 jets 	&  0.158  	&  0.036  	&  0.043  	&  $^{+ 0.002 }_{-0.001}$  	&  $^{+ 0.001 }_{-0.017}$  	\\
$\geq$ 4 / $\geq$ 3 jets 	&  0.207  	&  0.104  	&  0.117  	&  $^{+ 0.002 }_{-0.003}$  	&  $^{+ 0.031 }_{-0.001}$  	\\

\hline
\end{tabular}
}
\end{center}
\end{table}

 \begin{figure}
 {
   \centering
\includegraphics[width=\sizeFigTwo]{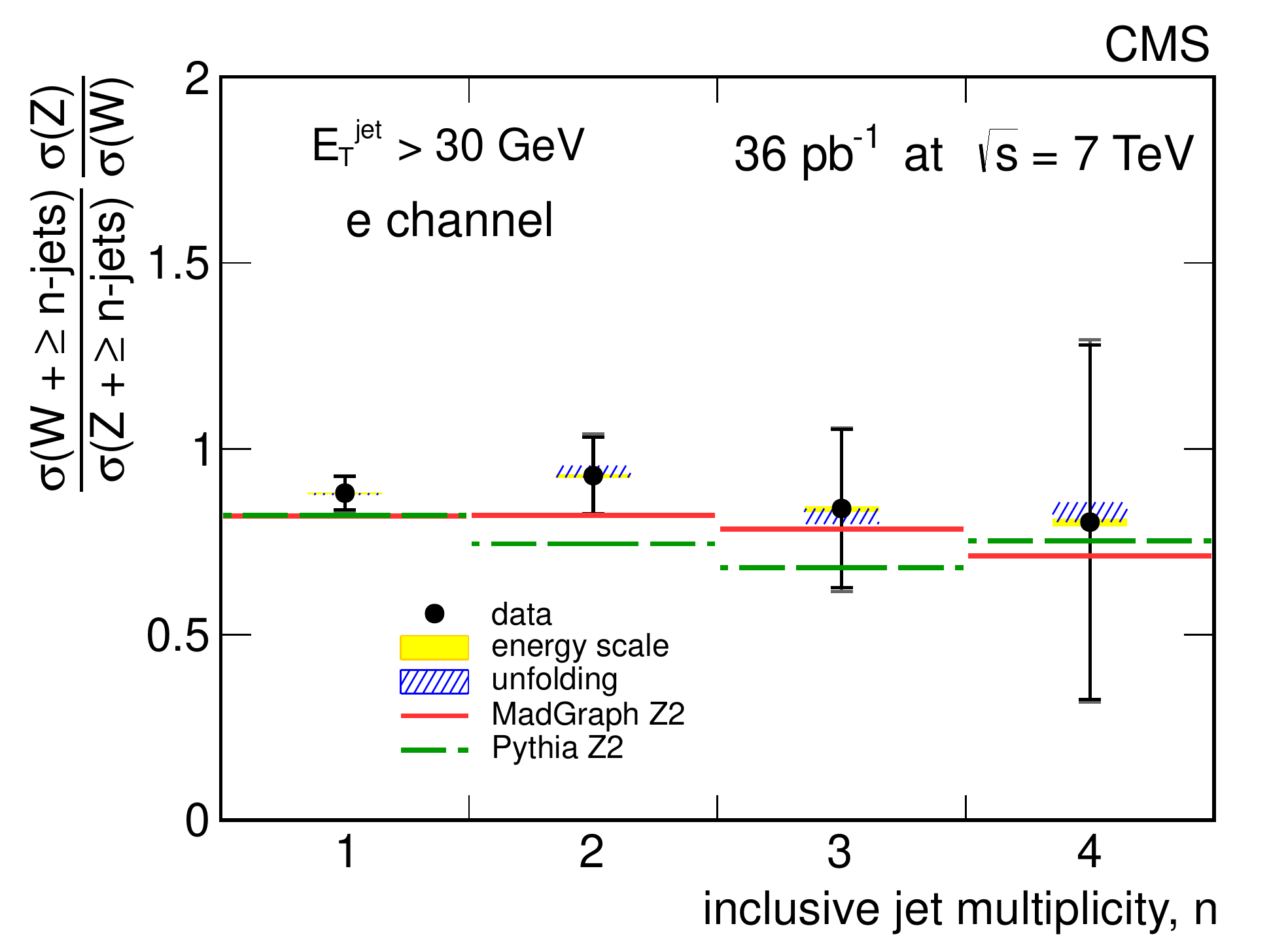}
\includegraphics[width=\sizeFigTwo]{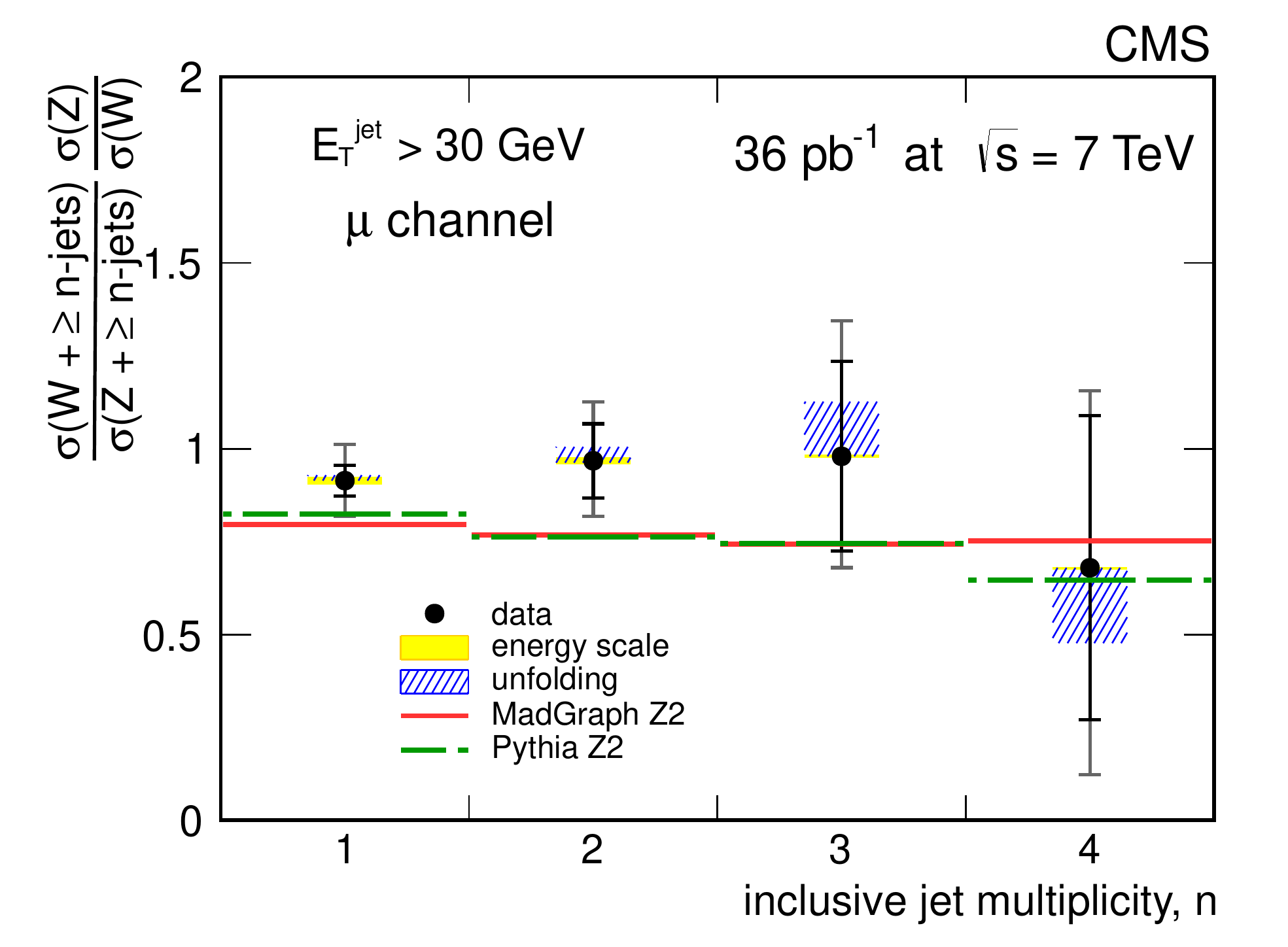}
   \caption{
 Ratio of the \wjets and \zjetsb cross sections for the
 electron channels (left) and the muon channels (right) as a function
 of the jet multiplicity. 
 The ratios are normalized to the
 inclusive W/Z cross section. The expectations from \MADGRAPH
 and \PYTHIA simulations, both with the Z2 tune, are shown. These expectations do not
 differ significantly and they are both in agreement with data.
 The difference between the expected values of the ratio in
 electron and muon channels is due to the larger electron acceptance in
 $\eta$.  Error bars on data are shown for the statistical and total uncertainties.
 The uncertainties due to the energy scale and the unfolding
 procedure are shown as yellow and hatched bands, respectively.
 \label{fig:WZratio}}
 }
 \end{figure}

The ratios of \wjets and \zjetsb cross sections are shown in Fig.~\ref{fig:WZratio}.
Many important systematic uncertainties, such as those on integrated luminosity
and jet energy scale, cancel in the ratio. The most significant
remaining systematic uncertainty is due to the
selection efficiency, which could be correlated between different bins because
of the \MT\ cut in the W candidate selection. The maximal difference observed
between the measured and expected values is at the level of one standard
deviation, neglecting the uncertainties on the theoretical
predictions. The difference between the expected value of the ratio in
the electron and muon channels is due to the larger electron acceptance in
$\eta$.

\begin{figure}
 {
   \centering
 \includegraphics[width=\sizeFigTwo]{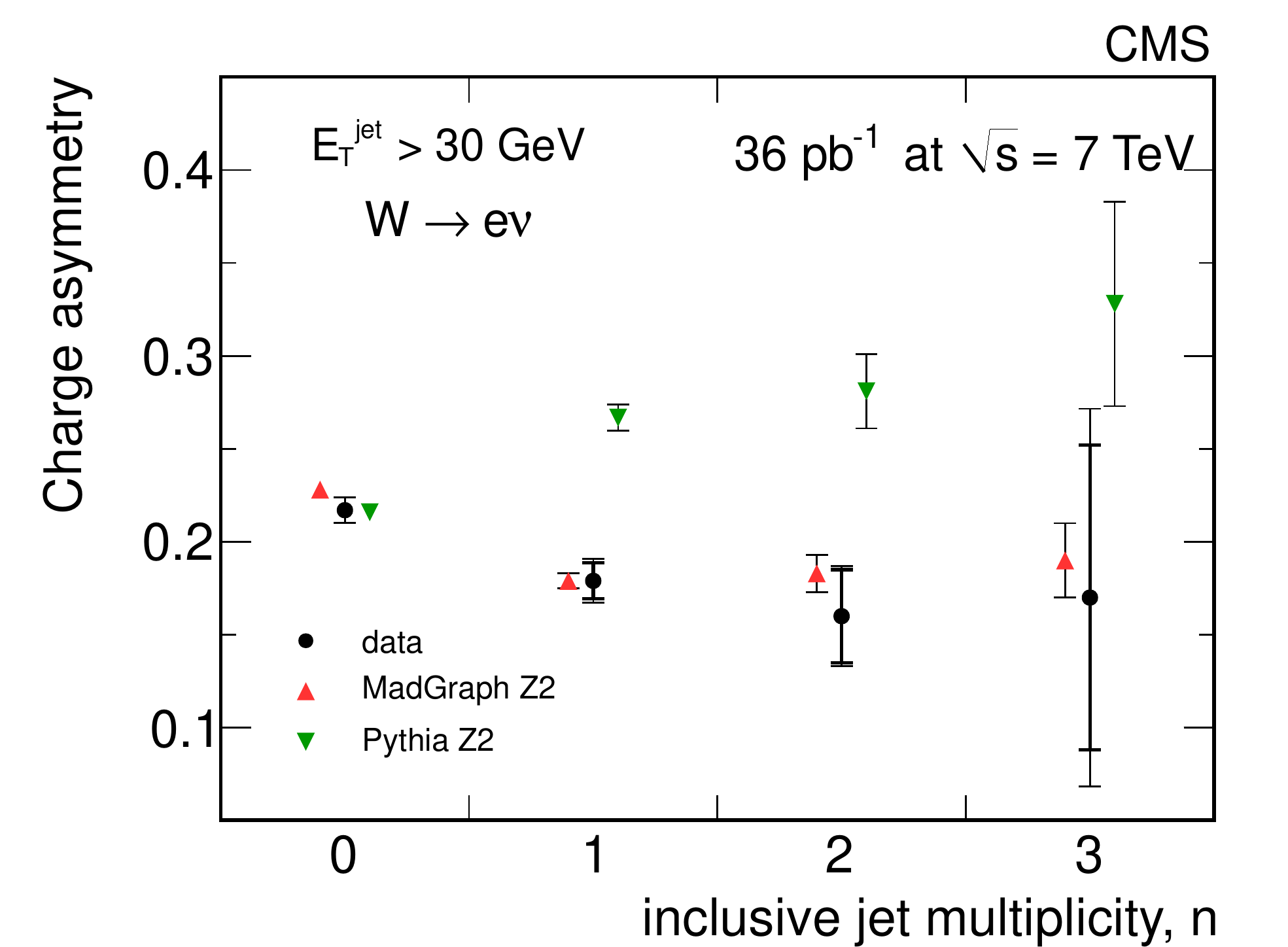}
 \includegraphics[width=\sizeFigTwo]{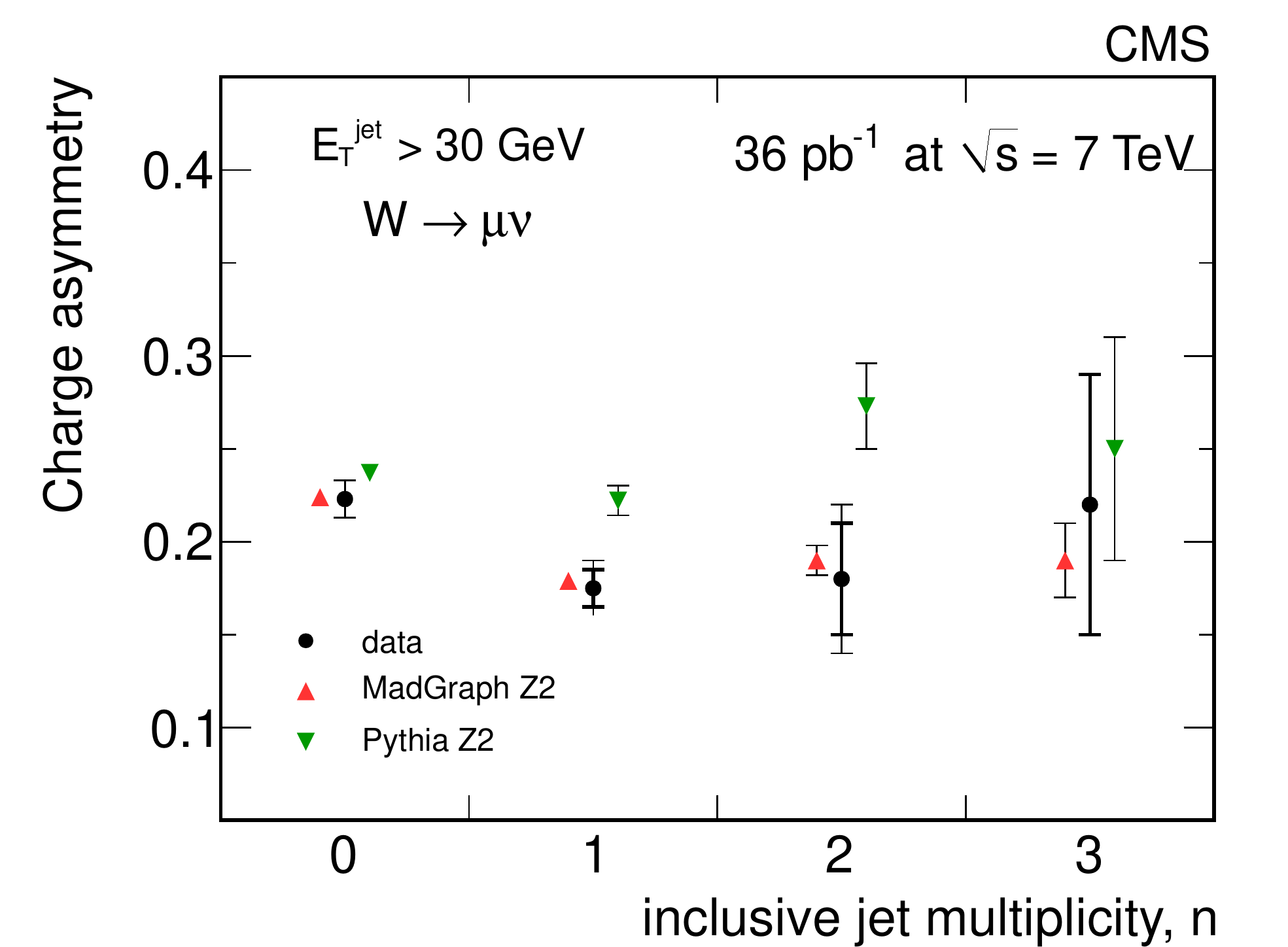}
   \caption{W charge asymmetry $\AW$ versus the inclusive jet multiplicity. Data are compared with predictions
     from the \MADGRAPH
     and \PYTHIA simulations. Error bars on the data points show the
     statistical and total uncertainties. Error bars on the simulation
     points correspond to the statistical uncertainty only. Left: electron
   decay channel, right: muon decay channel.
 \label{fig:WChargeRatios}}
 }
\end{figure}

The charge asymmetry, defined as $\AW = \frac{\sigma(\Wp)-\sigma(\Wm)}{\sigma(\Wp)+\sigma(\Wm)}$,
is measured as a function of the
number of jets in the events for the muon and electron channels, by
fitting separately the events with positive and negative lepton charge.
The sample size does not allow a measurement in events with four jets or more.
Table~\ref{tab:inclusiveWChargeRatios} and Fig.~\ref{fig:WChargeRatios}  show the measured value of the charge asymmetry
for the electron and muon events.
The systematic uncertainties include those from the jet energy scale, the charge misidentification,
and the positive versus negative lepton efficiency difference.
The results are compatible with the inclusive charge asymmetry measurement~\cite{Wasym}.

The charge asymmetry depends on the number of associated jets
because the fraction of u (d) quarks contributing to the process
is different in each case. The measured values are found to be in good agreement
with the predictions based on \MADGRAPH with the Z2 tune,
while \PYTHIA does not describe well the W charge asymmetry, even for events with a single associated jet.

\begin{table}
\caption{
W charge asymmetry $\AW$ in the data and in the \MADGRAPH and \PYTHIA (Z2 tune)
simulations for the electron and muon channel. The uncertainties on
the simulation values are statistical only.
\label{tab:inclusiveWChargeRatios}
}
\begin{center}
{
\begin{tabular}{ | c | c | c | c |}
\hline
 $n$ jets & data & \MADGRAPH Z2& \PYTHIA Z2\\
\hline\hline
\multicolumn{3}{|c|}{electron channel} \\
\hline
$\geq$ 0         &  $ 0.217\pm0.004\stat\pm0.006\syst $   & $ 0.228 \pm0.001  $   & $0.216\pm0.003$   \\
$\geq$ 1         &  $ 0.179\pm0.010\stat\pm0.007\syst $   & $ 0.179 \pm0.004 $    & $0.267\pm0.007$  \\
$\geq$ 2         &  $ 0.16\pm0.03\stat\pm0.01\syst $      & $ 0.183 \pm0.010 $    & $0.281\pm0.020$  \\
$\geq$ 3         &  $ 0.17 \pm0.08\stat\pm0.06\syst $     & $ 0.19  \pm0.02 $     & $0.33\pm0.05$   \\
\hline\hline
\multicolumn{3}{|c|}{muon channel} \\
\hline
$\geq$ 0         &  $0.223\pm0.003\stat\pm0.010\syst $       & $ 0.224\pm0.001 $   & $0.237\pm0.003$   \\
$\geq$ 1         &  $0.175\pm0.010\stat\pm0.011\syst $       & $ 0.179\pm0.003 $   & $0.222\pm0.008$    \\
$\geq$ 2         &  $ 0.18\pm0.03 \stat\pm0.02 \syst $       & $ 0.190\pm0.008 $   & $0.273\pm0.023$   \\
$\geq$ 3         &  $ 0.22\pm0.07 \stat\pm0.02 \syst $       & $ 0.19\pm0.02 $     & $0.26\pm0.06$   \\
\hline
\end{tabular}
}
\end{center}
\end{table}

\begin{figure}
  \centering
  \includegraphics[width=0.9\textwidth]{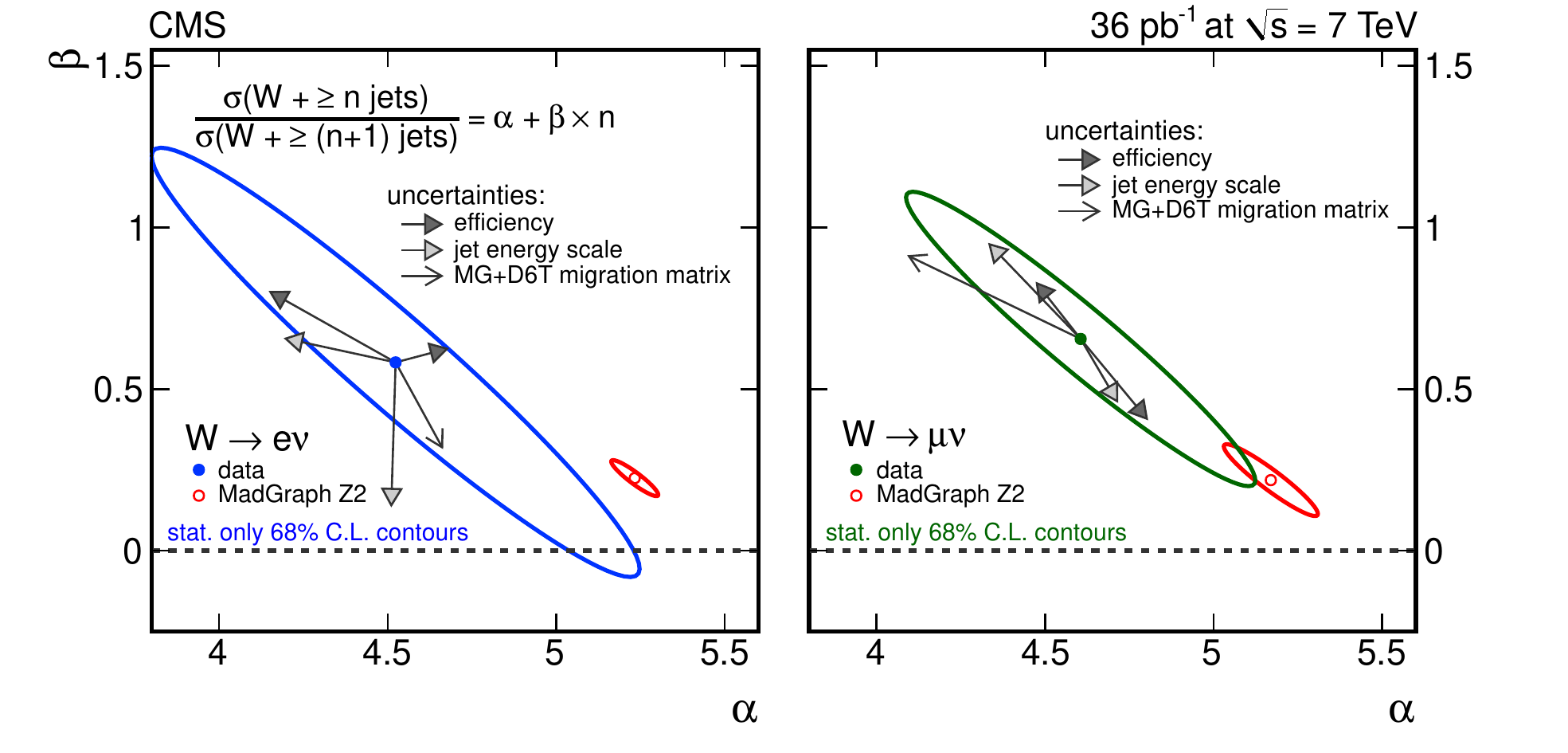}\\
  \includegraphics[width=0.9\textwidth]{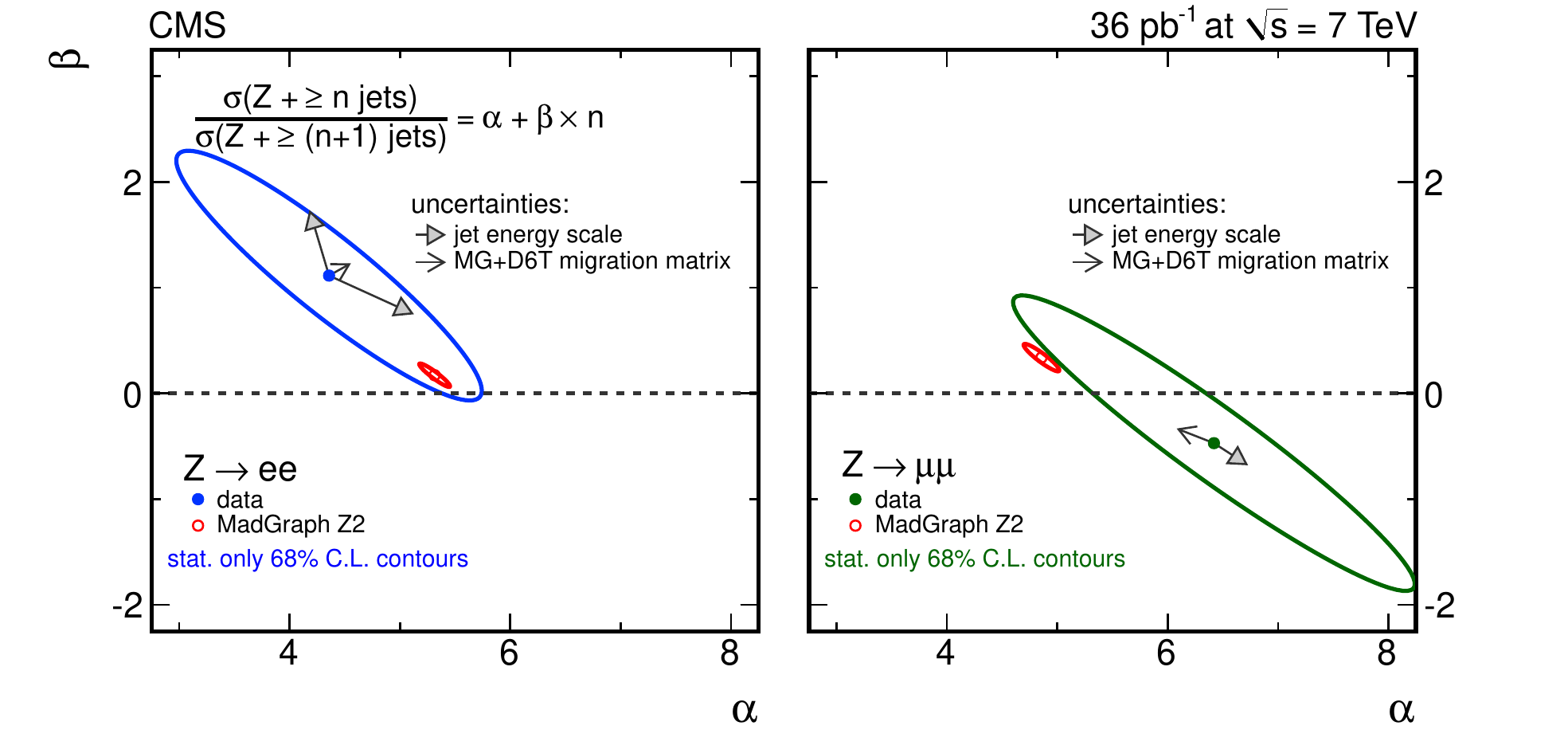}
  \caption{Fit results for the Berends--Giele scaling parameters
  $\alpha$ and $\beta$ after pileup subtraction, efficiency
  corrections, and unfolding of detector resolution effects: (top)
  \wjets, (bottom) \zjets, (left) electrons, (right) muons.
The data are compared with the expectations from the \MADGRAPH simulation with the Z2 tune.
The ellipses correspond to 68\% confidence level contours considering the
statistical uncertainty only, for both data and simulation.
The arrows show the displacement of
the central value when varying each indicated parameter by its
estimated uncertainty. The arrows labelled ``MG+D6T migration matrix''
correspond to the displacement when \MADGRAPH simulation with the D6T
tune is used for the unfolding.
    \label{fig:alphabeta}
    }
\end{figure}

Finally, we test for Berends--Giele scaling~\cite{BGscaling}
and measure its parameters. Events are
assigned to exclusive jet multiplicity bins (inclusive for
$n \ge 4$), and the corrected yields are fitted with the assumption
that they conform to a scaling function:
\begin{equation}
C_n \equiv \frac{\sigma_n}{\sigma_{n+1}} = \alpha \, ,
\label{eq:scaling}
\end{equation}
where $\sigma_n = \sigma(\PV+\ge n~{\mathrm{jets}})$, and $\alpha$ is a constant.
Previous measurements have shown that the Berends-Giele ratio can be approximately
constant~\cite{CDFW,D0W,D0ttbar}.
Phase-space effects, however, can violate this simple proportionality.
Therefore we introduce a second parameter, $\beta$,
to allow for a deviation from a constant scaling law:
\begin{equation}
\label{eq:BG}
    C_n=\alpha + \beta \, n .
\end{equation}
Because of the different production kinematics of the $n = 0$
sample, where no reconstructed jets recoil against
the vector boson, the scaling expressed in Eq.~(\ref{eq:BG})
is not expected to hold, so we do not include the $n = 0$ sample
in the fit.
The results of the fit for $\alpha$ and $\beta$ from the
\wjets and \zjetsb samples are shown in Fig.~\ref{fig:alphabeta}.
The values are given in the $(\alpha,\beta)$
plane and are compared with expectations from \MADGRAPH with
the Z2 tune for the underlying event description. Generator results are obtained at the particle jet
level for events within the lepton and jet acceptance of the selection.
The electron and muon expected
values differ mostly because of the $\Delta R>0.3$ requirement between the jets and
the leptons, which is applied only in the electron channel.
The ellipses correspond to 68\% confidence level contours using the
statistical uncertainty only.
The arrows show the displacement of the central value
when varying each indicated parameter by its estimated uncertainty.
The fit results are also reported in Tables~\ref{tab:alphabetaele} and
\ref{tab:alphabetamuons}. The data are in agreement with
expectations within one or two standard
deviations depending on the channel.
Furthermore, the Berends--Giele scaling hypothesis \ie, the relationship shown in
Eq.~(\ref{eq:scaling}), is confirmed to work well up to the production of
four jets. The $\beta$ parameter lies within one standard deviation from zero
for the \wjets case and within 0.5 standard deviations for the \zjetsb case.

\begin{table}[ht]
\centering
\caption{Results for the Berends--Giele parameters in the electron channel
compared with expectations from the \MADGRAPH Z2 simulation at the particle
level. The prediction uncertainty is statistical only.}
\label{tab:alphabetaele}
\begin{tabular}{ | c | c || c | c | c | c | c || c |}
\hline
 &          & data   & stat.      & energy scale   & efficiency    & tune & theory  \\
\hline
\hline
W & $\alpha$& $4.5$& $\pm 0.5$ & $_{-0.3}^{+0.1}$   & $_{-0.4}^{+0.2}$ & +0.1 & $5.20\pm0.05$  \\
  & $\beta$ & $0.6$& $\pm 0.4$ & $_{-0.4}^{+0.1}$   & $_{-0.1}^{+0.2}$ & -0.3 & $0.20\pm0.04$  \\
Z & $\alpha$& $4.4$& $\pm 0.9$ & $_{-0.2}^{+0.8}$   & $\pm 0.05$     & +0.2 & $5.3\pm0.1$  \\
  & $\beta$ & $1.1$& $\pm 0.8$ & $_{-0.4}^{+0.6}$   & $\pm 0.3$      & +0.1 & $0.17\pm0.07$  \\
\hline
\end{tabular}
\end{table}

\begin{table}[ht]
\centering
\caption{Results for the Berends--Giele parameters in the muon channel
compared with expectations from the \MADGRAPH Z2 simulation at the particle level. The
prediction uncertainty is statistical only.}
\label{tab:alphabetamuons}
\begin{tabular}{ | c | c || c | c | c | c | c || c |}
\hline
 & & data & stat. & energy scale & efficiency & tune & theory  \\
\hline
\hline
W & $\alpha$ & $4.6$  & $\pm 0.3$  & $^{+0.1}_{-0.3}$  & $^{+0.2}_{-0.1}$ & -0.5 & 5.17 $\pm$ 0.09  \\
  & $\beta$  &  $0.7$ & $\pm 0.3$  & $^{+0.3}_{-0.2}$  & $\pm 0.3$      & +0.3 & 0.22 $\pm$ 0.07  \\
Z & $\alpha$ & $6.4$  & $\pm 1.2$  & $^{+0.1}_{-0.3}$  & $\pm 0.1$      & -0.3 & 4.8  $\pm$ 0.1    \\
  & $\beta$  & $-0.5$ & $\pm 0.9$  & $^{+0.1}_{-0.2}$  & $\pm 0.2$      & +0.1  & 0.34 $\pm$ 0.09  \\
\hline
\end{tabular}
\end{table}

\section{Summary}
\label{sec:conclusions}
\par
The rate of jet production in association
with a \W or \Z boson was measured in pp collisions
at $\sqrt{s} = 7$~TeV.  The data were collected with the
CMS detector in 2010 and correspond to an integrated
luminosity of $36 \pbi$.  The \wjets and \zjetsb
samples were reconstructed in the electron and muon
decay channels.  The  measurement was performed using particle-flow jets with
$\ET > 30$~GeV and clustered with the anti-$k_T$ algorithm
with a size parameter $R = 0.5$.
\par
The \ET spectra agree well with
the predictions from simulations
based on \MADGRAPH interfaced with \PYTHIA, and using the Z2 tune for the underlying
event description.
\par
Detector resolution effects were unfolded to extract the exclusive jet multiplicity distributions
and to measure the ratios of the normalized inclusive rates
$\sigma(\PV+\ge n~{\mathrm{jets}})/\sigma(\PV)$.
The ratio of the W + jets to Z + jets cross sections,
$[\sigma(\PW+\ge n~{\mathrm{jets}})/\sigma(\PZ+\ge n~{\mathrm{jets}})]/[\sigma(\PZ)/\sigma(\PW)]$,
and the W charge asymmetry, $\AW$, were also measured
as functions of the jet multiplicity.
Finally, a quantitative test of Berends--Giele scaling,
parametrized as a function of two parameters determined by
a fit, was performed.
\par
All results are in agreement with the predictions of the \MADGRAPH
generator, using matrix-element calculations for final states
with jets, matched with \PYTHIA parton shower.
In contrast, the simulation based on parton showers alone fails to describe the
jet rates for more than one jet, and the W charge asymmetry, even in
the case of only one jet.

\section*{Acknowledgements}

\hyphenation{Bundes-ministerium Forschungs-gemeinschaft
  Forschungs-zentren} We wish to congratulate our colleagues in the
CERN accelerator departments for the excellent performance of the LHC
machine. We thank the technical and administrative staff at CERN and
other CMS institutes. This work was supported by the Austrian Federal
Ministry of Science and Research; the Belgium Fonds de la Recherche
Scientifique, and Fonds voor Wetenschappelijk Onderzoek; the Brazilian
Funding Agencies (CNPq, CAPES, FAPERJ, and FAPESP); the Bulgarian
Ministry of Education and Science; CERN; the Chinese Academy of
Sciences, Ministry of Science and Technology, and National Natural
Science Foundation of China; the Colombian Funding Agency
(COLCIENCIAS); the Croatian Ministry of Science, Education and Sport;
the Research Promotion Foundation, Cyprus; the Estonian Academy of
Sciences and NICPB; the Academy of Finland, Finnish Ministry of
Education and Culture, and Helsinki Institute of Physics; the Institut
National de Physique Nucl\'eaire et de Physique des Particules~/~CNRS,
and Commissariat \`a l'\'Energie Atomique et aux \'Energies
Alternatives~/~CEA, France; the Bundesministerium f\"ur Bildung und
Forschung, Deutsche Forschungsgemeinschaft, and Helmholtz-Gemeinschaft
Deutscher Forschungszentren, Germany; the General Secretariat for
Research and Technology, Greece; the National Scientific Research
Foundation, and National Office for Research and Technology, Hungary;
the Department of Atomic Energy and the Department of Science and
Technology, India; the Institute for Studies in Theoretical Physics
and Mathematics, Iran; the Science Foundation, Ireland; the Istituto
Nazionale di Fisica Nucleare, Italy; the Korean Ministry of Education,
Science and Technology and the World Class University program of NRF,
Korea; the Lithuanian Academy of Sciences; the Mexican Funding
Agencies (CINVESTAV, CONACYT, SEP, and UASLP-FAI); the Ministry of
Science and Innovation, New Zealand; the Pakistan Atomic Energy
Commission; the State Commission for Scientific Research, Poland; the
Funda\c{c}\~ao para a Ci\^encia e a Tecnologia, Portugal; JINR
(Armenia, Belarus, Georgia, Ukraine, Uzbekistan); the Ministry of
Science and Technologies of the Russian Federation, the Russian
Ministry of Atomic Energy and the Russian Foundation for Basic
Research; the Ministry of Science and Technological Development of
Serbia; the Ministerio de Ciencia e Innovaci\'on, and Programa
Consolider-Ingenio 2010, Spain; the Swiss Funding Agencies (ETH Board,
ETH Zurich, PSI, SNF, UniZH, Canton Zurich, and SER); the National
Science Council, Taipei; the Scientific and Technical Research Council
of Turkey, and Turkish Atomic Energy Authority; the Science and
Technology Facilities Council, UK; the US Department of Energy, and
the US National Science Foundation.

Individuals have received support from the Marie-Curie programme and
the European Research Council (European Union); the Leventis
Foundation; the A. P. Sloan Foundation; the Alexander von Humboldt
Foundation; the Belgian Federal Science Policy Office; the Fonds pour
la Formation \`a la Recherche dans l'Industrie et dans l'Agriculture
(FRIA-Belgium); the Agentschap voor Innovatie door Wetenschap en
Technologie (IWT-Belgium); and the Council of Science and Industrial
Research, India.

\bibliography{auto_generated}   

\cleardoublepage \appendix\section{The CMS Collaboration \label{app:collab}}\begin{sloppypar}\hyphenpenalty=5000\widowpenalty=500\clubpenalty=5000\textbf{Yerevan Physics Institute,  Yerevan,  Armenia}\\*[0pt]
S.~Chatrchyan, V.~Khachatryan, A.M.~Sirunyan, A.~Tumasyan
\vskip\cmsinstskip
\textbf{Institut f\"{u}r Hochenergiephysik der OeAW,  Wien,  Austria}\\*[0pt]
W.~Adam, T.~Bergauer, M.~Dragicevic, J.~Er\"{o}, C.~Fabjan, M.~Friedl, R.~Fr\"{u}hwirth, V.M.~Ghete, J.~Hammer\cmsAuthorMark{1}, S.~H\"{a}nsel, M.~Hoch, N.~H\"{o}rmann, J.~Hrubec, M.~Jeitler, W.~Kiesenhofer, M.~Krammer, D.~Liko, I.~Mikulec, M.~Pernicka, B.~Rahbaran, H.~Rohringer, R.~Sch\"{o}fbeck, J.~Strauss, A.~Taurok, F.~Teischinger, C.~Trauner, P.~Wagner, W.~Waltenberger, G.~Walzel, E.~Widl, C.-E.~Wulz
% ...
\vskip\cmsinstskip
% ...
1:~~Also at CERN, European Organization for Nuclear Research, Geneva, Switzerland\\
% ...

\end{sloppypar}
\end{document}